\documentclass{aa}

\usepackage{epsfig,graphicx,natbib}
\usepackage{longtable}
\usepackage{amssymb}
\usepackage{amsmath}
\usepackage{mathrsfs} 
\usepackage{color}
\usepackage{subcaption}
\usepackage{booktabs}
\usepackage{xspace}
\usepackage{float}
\usepackage{longtable}
\usepackage{breqn}
\usepackage{algpseudocode}
\usepackage{algorithm}
\usepackage{hyperref}
\usepackage[raggedright]{sidecap}

\usepackage{todonotes}

\newcommand{\norm}[1]{\lVert #1 \rVert}

\renewcommand{\textit}[1]{#1}

\begin{document}

\title{Accelerating the CLEAN algorithm of radio interferometry with convex optimization}

\author{Hendrik M\"uller \inst{1,2,*} \and 
        Mingyu Hsieh \inst{2} \and
        Sanjay Bhatnagar \inst{2}}

\institute{\inst{1}Max-Planck-Institut für Radioastronomie, Auf dem Hügel 69, D-53121 Bonn, Germany\\
\inst{2}National Radio Astronomy Observatory, P.O. Box O, Socorro, NM 87801, USA\\
\inst{*}\email{hmuller@nrao.edu}\\
}

\date {Received  / Accepted}

\authorrunning{M\"uller}
\titlerunning{Accelerating CLEAN with convex optimization}

\abstract
%context heading (optional)
{In radio-interferometry, we recover an image from an incompletely sampled Fourier data. The de-facto standard algorithm, the Cotton-Schwab CLEAN, is iteratively switching between computing a deconvolution (minor loop) and subtracting the model from the visibilities (major loop).}
% aims heading (mandatory)
{The next generation of radio interferometers is expected to deal with much higher data rates, image sizes and sensitivity, making an acceleration of current data processing algorithms necessary. We aim to achieve this by evaluating the potential of various well-known acceleration techniques in convex optimization to the major loop. For the present manuscript, we limit the scope to study these techniques only in the CLEAN framework.}
% methods heading (mandatory)
{To this end, we identify CLEAN with a Newton scheme, and use this chain of arguments backwards to express Nesterov acceleration and conjugate gradient orthogonalization in the major and minor loop framework.}
% results heading (mandatory)
{The resulting algorithms are simple extensions of the traditional framework, but converge multiple times faster than traditional techniques, and reduce the residual significantly deeper. These improvements achieved by accelerating the major loop are competitive to well-known improvements by replacing the minor loop with more advanced algorithms, but at lower numerical cost. The best performance is achieved by combining these two developments.}
% conclusions heading (optional)
{CLEAN remains among the fastest and most robust algorithms for imaging in radio interferometry, and can be easily extended to an almost an order of magnitude faster convergence speed and dynamic range. The procedure outlined in this manuscript is relatively straightforward and could be easily extended.}

\maketitle

\section{Introduction}
The observable in a radio interferometer are the correlated (and time-shifted) signals recorded at the individual antennas. These are approximately related to the on-sky emission through a Fourier transform as described by the van-Cittert-Zernike theorem \citep{Thompson2017}. For the data processing in radio interferometric imaging, an image must be reconstructed from sparsely sampled (on a non-equidistant grid), noisy Fourier coefficients (called visibilities). Equivalently, this problem is reformulated as a deconvolution problem in which the residual map (inverse Fourier transform of the visibilities) needs to be deconvolved with the beam (inverse Fourier transform of the sampling pattern, also called the point spread function).

Historically the imaging problem has been solved by the CLEAN algorithm \citep{Hogbom1974}, a matching pursuit algorithm in which the image is inversely modeled by the subtraction of individual model components from the residual. Iteratively the algorithm searches for the maximum in the current residual, shifts the beam to that location and subtracts a fraction of the beam. In this way, the emission is modeled iteratively by delta components. The CLEAN algorithm has been further developed and modernized into multiscalar and multifrequency variants \citep{Bhatnagar2004, Cornwell2008, Rau2011, Offringa2014, Offringa2017, Mueller2023a, Mueller2025}. In its most common implementation, as for example available from the CASA package \citep{CASA2022}, the CLEAN algorithm operates in the image domain. We refer to this iterative procedure to deconvolution as the minor loop. Data consistency is ensured by an external loop (referred to as the major loop). The residual is updated by Fourier transforming and degridding the current image model, comparing with the observed visibilities, and subsequently gridding and inverse Fourier transforming the difference of the visibilities. As first proposed by \citet{Schwab1984}, one typically alternates between these two steps (computing the deconvolution and updating the residual). Since these two steps are inherently linked to each other, we will refer to the combination of major loop and minor loop as the CLEAN algorithm for the remainder of this manuscript.

A wide range of alternative imaging approaches has been suggested in the past. Historically, this includes the Maximum-Entropy Method \citep{Cornwell1985}, regularized maximum-likelihood \citep[e.g.][]{Akiyama2017, Chael2018, Mueller2023c, Mus2024c}, compressive sensing \citep[e.g.][]{Starck1994, Wiaux2009, Garsden2015, Onose2017, Mueller2022, Mueller2023b}, Bayesian approaches \citep[e.g.][]{Junklewitz2016, Broderick2020b, Arras2021} and recently AI-based ideas \citep[e.g.][]{Connor2022, Sun2022, Aghabiglou2024}. Most of these approaches formulate the problem as a forward modelling problem. Forward modelling is concerned with describing the process by which the sought-after signal is mapped and solving it in a statistically consistent manner. We use the frequently used terminology of forward modelling here to highlight the difference between the strategy of including the instrumental response directly in the forward operator, opposed to correcting inversely for instrumental, calibration and gridding effects, and projecting them into the image domain (the strategy applied by most CLEAN implementation through the major and minor loop). However, the terminology should not confuse that also the CLEAN algorithm is a forwardly modelling algorithm that ensures data compatibility through the major loop.

These techniques address several key limitations of the CLEAN approach. Unlike CLEAN, which implicitly encodes regularization within the algorithm and by user-based choices, forward-modeling methods optimize well-defined objective functions. They also provide a natural framework for joint calibration and deconvolution, adapting also well to cases where the forward operator becomes non-linear, and they can incorporate uncertainty quantification. Moreover, they resolve the inherent dichotomy in CLEAN between the model (a set of CLEAN components) and the final image (components convolved with the clean beam). The forward formulation further allows the inclusion of more sophisticated prior information than assuming a sparse set of CLEAN components, for example through prior distributions, explicit regularizers, or learned/inferred features. As a result, several science cases have demonstrated that forward-modeling approaches can outperform CLEAN-based methods in terms of dynamic range and resolution \citep[e.g.][]{Arras2021, Dabbech2024}, although opposite findings have also been reported \citep{Offringa2017, Bester2025}. A notable area where CLEAN is increasingly being supplanted by forward-modeling techniques is global millimeter-wavelength Very Long Baseline Interferometry (VLBI), where the limited number of antennas and modest signal-to-noise ratios pose particular challenges. Nonetheless, CLEAN remains applicable and widely used in such settings as well \citep[e.g.][]{eht2019d, eht2021a, eht2022c, Lu2023, eht2024b}.

Despite these developments, CLEAN remains the standard method in radio interferometry. This is mainly because of its maturity, robustness, adaptivity, and its speed. The speed of CLEAN is achieved by its rather simple deconvolution procedure mostly including steps of only linear numerical complexity not scaling with the number of visibilities, and the splitting of the data analysis problem into the major and minor loop. Noteworthy, latter idea has been recently adapted to accelerate Bayesian imaging \citep{Roth2024} and AI-based imaging \citep{Aghabiglou2024} as well, achieving up to an order of magnitude in speed-up. 

Alternative approaches, despite suggestive performance and precision, have not yet enjoyed wide-spread application on an observatory level, partly caused by the significant development work that would still be necessary to mature these approaches into application-ready, automated pipelines, robustly adapting to all observing modes, science cases and source structures accessible to an instrument.  Moreover, also CLEAN implementations have seen substantial additional development work in recent years \citep{Hsieh2021, Hsieh2022b, Mueller2023a, Mueller2025, Jarret2024}, mimicking some of the performance gains claimed by alternative approaches.

A new generation of radio interferometers is either under development or actively planned, including the Square Kilometre Array \citep[SKA][]{Dewdney2009}, the next-generation Very Large Array \citep[ngVLA][]{Murphy2018, Selina2018}, and the DSA-2000 \citep{Hallinan2019}. Additionally, the Atacama Large Millimeter/submillimeter Array (ALMA) is slated for a significant enhancement through its Wideband Sensitivity Upgrade \citep[WSU][]{Carpenter2022}. These next-generation instruments will operate at substantially higher data rates and with markedly improved sensitivity. These advancements introduce considerable challenges for data processing. In particular, all imaging and calibration techniques must scale effectively to accommodate the data volumes produced by these arrays. Furthermore, standard simplifying assumptions, such as representing the sky brightness distribution using point source models, may no longer be valid. This necessitates the development and deployment of more computationally intensive and sophisticated algorithms. 

It has been demonstrated that the driving numerical cost may be the gridding \citep{ngvla_comp_4, ngvla_comp_7}. This indeed questions the use of many forward modeling approaches, albeit their superior dynamic range demonstrated in a few test cases and underpins the recent efforts to recast them in an efficient minor loop/major loop skeleton. Scalability is the paramount priority and requirement for the construction of next generation data processing pipelines \citep{ngvla_comp_7, Kepley2023, Hunter2023, Bhatnagar2025}. Improved resolution and dynamic range would benefit many science cases, but the majority of observational data, especially the large mosaics and surveys that dominate numerical cost in observatory-driven automatized pipelines, remain well served by CLEAN variants \citep[see e.g. the discussions in][]{Kepley2023, Bester2025}. As a consequence, multiscalar variants of CLEAN, shown to be performing well, fast and robust, remain among the first choices for constructing pipelines for these arrays and are under active development and deployment by observatories \citep[e.g.][]{Hsieh2021, Hsieh2022a, CASA2022, Hunter2023, Bhatnagar2025}.

Despite this remarkable longevity, CLEAN operates a relatively simple optimization frame, not making use of accelerating techniques that became popular within the last two decades. Regularized Maximum Likelihood and compressive sensing that formulate the problem directly as an optimization problem have developed and utilized more advanced settings. In this manuscript, we aim to combine these two approaches, i.e. we apply concepts of accelerating numerical optimization to the classical CLEAN framework. As a result, we construct algorithms that are still based in the robust and proven framework of CLEAN, but converge faster in terms of number of major loop iterations, significantly reducing computational time in consequence.

This will be done by first formulating CLEAN in the common language of optimization and then, in the opposite logical direction, deriving CLEAN alternatives in the context of accelerated optimization algorithms. The manuscript is structured along this logic. We will first in Sec. \ref{sec:CLEAN_opt} discuss how we can understand CLEAN as an optimization technique, and then backwards in Sec. \ref{sec:opt_CLEAN} how to formulate the optimization techniques in the CLEAN frameworks. We show results on synthetic and real data applications in Sec. \ref{sec:validation} and \ref{sec:results}, and present our conclusions in Sec. \ref{sec:conclusions}.

\section{CLEAN as an optimization technique} \label{sec:CLEAN_opt}

\subsection{Radio Interferometry}
The correlated signal between two antennas is approximately the Fourier transform of the on-sky emission $I(l,m)$:
\begin{align}
\mathcal{V} (u, v) = \int \int I(l, m) e^{-2 \pi i (l u + m v)} dl dm. \label{eq: vis}
\end{align}
Here $\mathcal{V}$ denotes the visibilities (the measurements), $u,v$ are harmonic coordinates and $l,m$ direction cosines on-sky. In practice, only a sparse subset of the visibilities is measured, the visibilities are corrupted by thermal noise and station-based and direction-dependent calibration effects. All of these operations are summarized in the Radio Interferometry Measurement Equation \citep{Hamaker1996, Smirnov2011}:
\begin{align} \nonumber
V_{ij}(t) &= G_i(t)\! \left( \int_{\Omega} E_i(\hat{s},t)\,I(\hat{s},t)\,E_j^\dagger(\hat{s},t)\,e^{-2\pi i\,u_{ij}\cdot\hat{s}}\,\mathrm{d}\Omega\right)G_j^\dagger(t)\\
&+ \epsilon_{ij}(t).
\label{eq:rime-cont}
\end{align}
Here $G_i$ and $G_j$ describe direction-independent gains, $E_i$ and $E_j$ direction-dependent gains, and $\epsilon$ additive thermal noise.

An important numerical complexity stems from the fact that the visibilities are not taken on an equidistant grid. Computing the residual and the dirty beam therefore includes more operations than the simple application of the Fast Fourier Transform (FFT). We first have to convolve the measured visibilities in the Fourier domain with a gridding kernel before applying an inverse FFT, and then correct for this convolution in the image domain (normalization). This procedure is typically referred to as the gridding.

For the remainder of the manuscript, we will denote the discretized version of the measurement operator by $\Phi$ and study the problem:
\begin{align}
    V = \Phi I + \epsilon. \label{eq: forward_general}
\end{align}
We refer the interested reader to \citet{Bester2025, Bhatnagar2025} for a detailed derivation of the matrix form of $\Phi$. The measurement operator $\Phi$ is a rank-deficient matrix. We will refer to the transposed matrix of $\Phi$ by $\Phi^T$ throughout this manuscript.

The data fidelity between a current guess solution $I_k$ and the observed visibilities is described by the $\chi^2$-metric:
\begin{align}
    \chi^2(V,I_k) = \sum_i \frac{\norm{(V(u_i,v_i)-\left(\Phi I_k\right)(u_i,v_i)}^2}{\sigma_i^2},
\end{align}
where the index $i$ runs over all observed visibility points and $\sigma$ is the associated thermal noise level. In fact, this data fidelity would correspond to natural weighting. It is common practice to re-weight the visibility data points to enhance specific spatial features. For example, the uniform weighting of all data points leads to a higher resolution at the cost of sensitivity. A detailed description of common weighting schemes and their effect is provided by \citet{Briggs1995}. We will not dive deeper into this topic here, and rather assume that all weightings and tapers applied to the data are fully described by a fixed array of weights $\sigma$.

In the following, we will denote the domain of the operator $\Phi$ by $\mathbf{X}$ (i.e. the image plane), and its co-domain by $\mathbf{Y}$, i.e.:
\begin{align}
    \Phi: \mathbf{X} \rightarrow \mathbf{Y},
\end{align}
and equip the codomain with the topology described by the Gram matrix $G_{\mathbf{Y}} = \mathrm{diag}{1/\sigma_i^2}$. With this topology, we can define the $\chi^2$ objective functional as:
\begin{align}
    J: \theta \mapsto \frac{1}{2}\norm{V-\Phi \theta}^2_{\mathbf{Y}}.
\end{align}
Since for an arbitrary vector $r$, we have $\norm{r}_{\mathbf{Y}}^2 = \langle r, r\rangle_{\mathbf{Y}} = r^+ G_{\mathbf{Y}} r$, it is easy to show that $J(\theta) = 1/2 \cdot \chi^2(V,\theta)$ holds. We will denote the adjoint $\Phi$ with respect to that topology by $\Phi^+ = \Phi^T G_{\mathbf{Y}}$.

\subsection{Cotton Schwab Algorithm}
The so-called Cotton-Schwab algorithm \citep{Schwab1984} splits the imaging problem into a major loop and minor loop. It is the de-facto standard algorithm used in radio interferometry. We will refer to this set-up simply as CLEAN in the remainder of this manuscript.

A major loop iteration is the operation to update the current residual with the current guess solution $I_k$ at iteration $k$, i.e. degridding the model, comparing it to the observed visibilities, and gridding, and evaluating the inverse FFT on the difference. In the notation of the operators $\Phi$ and $\Phi^+$ introduced in the previous subsection, this may be formulated by the operation:
\begin{align}
    I_k \mapsto \Phi^+ \left( V - \Phi I_k \right).
\end{align}

In the minor loop, we aim to model the image structure. We reformulate Eq. \eqref{eq: forward_general} as a deconvolution problem:
\begin{align}
    I^{res} = B * I_k + N, \label{eq: deconvolution}
\end{align}
where $I^{res} := \Phi^+ V$, $B := \Phi^+ \Phi$ and $N := \Phi^+ \epsilon$. Here, $I^{res}$ is called the residual, and $B$ the beam (the inverse Fourier transform of the sampling pattern). The beam (also called the point spread function) can also be interpreted as the response to a point source. We will use these notations for the remainder of this manuscript.

A minor loop iteration consists of the substeps of finding the peak in the current residual, shift the beam to that position, subtract a fraction of the shifted beam from the residual, and store the position and strengths of the component in a list of delta components. Looping over these operations until convergence is achieved, we effectively solve the deconvolution problem Eq. \eqref{eq: deconvolution}. We will refer to the minor loop iteration in the following as $p_{k} = \mathrm{minorloop}(I_{k}^{res}, B)$, where $p_k$ is the model at iteration $k$, $I_k^{res}$ the residual at iteration $k$ and $B$ the convolution kernel.

An outline of the resulting algorithm is shown in Tab. \ref{tab: clean}. We iteratively solve the deconvolution problem by a minor loop, and then update the residual (major loop). The final model is $\theta_k$. It is common to convolve the model with the so-called clean beam (a Gaussian fit to the main lobe of the dirty beam), and add the last residual to account for any non-captured flux.

\begin{algorithm}
\caption{Classical CLEAN}

\begin{algorithmic}
\State $\theta_0 = 0$, $I_0^{res} = \Phi^+V$, $B = \Phi^+\Phi \delta$
\State $p_0 = \mathrm{minorloop}(I_0^{res}, B)$
\While{$k \mapsto k+1$ until convergence}
\State $\theta_{k+1} = \theta_k+p_k$
\State Update residual $I_{k+1}^{res}$ with new guess solution $\theta_{k+1}$
\State $p_{k+1} = \mathrm{minorloop}(I_{k+1}^{res}, B)$
\EndWhile
\end{algorithmic}

\label{tab: clean}
\end{algorithm}

\subsection{CLEAN as a Newton Algorithm} \label{sec: CLEAN_Newton}

The minor loop is target of many algorithmic improvements, most prominently in the context of MS-CLEAN variants, e.g. \citep{Bhatnagar2004, Cornwell2008, Rau2011, Offringa2014, Offringa2017, Hsieh2021, Hsieh2022b, Mueller2023a, Mueller2025}. In this manuscript, we however do not focus on the exact version of the minor loop, but rather focus on algorithmic accelerations possible at the algorithm flow of the major loop iterations. To this end, we will first interpret CLEAN as an optimization technique. It should be noted that this identification is not a mathematically rigorous proof. We are rather interested in identifying broad similarities and ideas that we can transfer between frameworks.

CLEAN is a matching pursuit approach, probably closest related to a Least Absolute Shrinkage and Selection Operator (LASSO) problem, see e.g. the recent discussion in \citet{Jarret2024}. However, CLEAN is not equivalent to the LASSO problem, and solves the linear problem in an iterative non-linear way. We will not discuss this point further, just assume that computing a minor loop either with CLEAN or MS-CLEAN is our preferred way of solving the deconvolution problem Eq. \eqref{eq: deconvolution}.

In the following, we want to minimize the data fidelity functional $J$. So we need to compute the gradient and the Hessian. The gradient of $J$ at the location of a guess solution $\theta$ is:
\begin{align}
    \nabla J (\theta) = \Phi^+ (V - \Phi \theta) = I^{res}(\theta). \label{eq: gradient}
\end{align}
The Hessian will be:
\begin{align}
    \nabla^2 J(\theta) = \Phi^+ \Phi = \left( \theta \mapsto B * \theta \right).
\end{align}
We see that the gradient of the objective functional is the update step of the residual, i.e. a single major loop iteration. The Hessian corresponds to the convolution with the point spread function. These findings have been reported in radio interferometry for a long time and are recently discussed in detail in \citet{Bhatnagar2025, Bester2025}. We are working on the identification of the gradient step with the major loop and the inversion of the Hessian as the minor loop for the construction of new algorithms in Sec. \ref{sec:opt_CLEAN}.

Finally, we would like to point out similarities between CLEAN and Newton minimization as a motivation for the construction of novel algorithms. One of the most famous optimization algorithms is the Newton algorithm. Newton minimization to minimize an objective $J$ would consist of the two step procedure:
\begin{enumerate}
    \item Find a solution $p$ solving $\nabla^2 J (\theta) p = \nabla J (\theta) $
    \item Update guess solution $\theta \leftarrow \theta - p$, update gradient $\nabla J(\theta)$ with new model, proceed with step 1
\end{enumerate}
Plugging in the gradient and the Hessian, we see that the first step corresponds to solving the deconvolution problem Eq. \ref{eq: deconvolution}, i.e. we perform the minor loop. The second step computes essentially the update of the residual, i.e. a major loop iteration.

In this way, we can identify CLEAN with Newton minimization of the data fidelity to the observed visibilities. We would like to stress here that this correspondence is not mathematically correct, we rather just point out similarities. We have not included the regularization term that would describe CLEAN in this description (and rather just replaced the inversion of the beam with the minor loop), and the Newton algorithm actually assumes that the beam is invertible, which is not the case for radio-interferometry. For the scope of this manuscript however, this erroneous correspondence may be sufficient to motivate alternative CLEAN major loop frameworks.

\section{Optimization Techniques as CLEAN} \label{sec:opt_CLEAN}
In the previous section, we started with CLEAN, and tried to express it as an optimization tool, seeing similarities in philosophy (albeit not mathematical equivalence) to Newton minimization of the objective functional $J$. The similarities exploit the well-known finding that the gradient of $J$ is calculated within major loop iterations, and the minor loop is one way to undo the convolution with the point spread function, the Hessian of $J$.

In this manuscript, we are interested in frameworks to accelerate the convergence in terms of major loop iterations, still rooted in the robust and fast CLEAN perspective. To this end, we propose to apply the reverse logic of the last section to derive algorithms. We will start from an optimization algorithm that may be well suited to solve the problem adequately, and then we try to build a correspondence to the CLEAN framework, by replacing gradient steps with the major loop and inversion of the Hessian with minor loop iterations.

\subsection{Implicit Gradient Descent} \label{sec: implicit_gd}
Let us first start with some simple instructive example showcasing the logic behind this work. However, the derived algorithm may have only limited potential. To this end, we study the implicit gradient descent algorithm. Simple gradient descent would read:
\begin{align}
    \theta_{i+1} = \theta_{i} - \alpha \nabla J(\theta_i),
\end{align}
where $\alpha$ is a real valued, positive step-size parameter. Plain gradient descent, would not be very useful in aperture synthesis. We would essentially just add residuals here (i.e. only perform major loop iterations without intermediate minor loop cleaning). An interesting alternative is the implicit gradient descent algorithm for which the gradient is evaluated at the new position rather than the old one. In some instances, implicit gradient descent performs better than usual gradient descent:
\begin{align}
    \theta_{i+1} = \theta_{i} - \alpha \nabla J(\theta_{i+1}).
\end{align}
With Eq. \eqref{eq: gradient}, we get: 
\begin{align}
    \theta_{i+1} = \theta_i - \alpha \Phi^+ V + \alpha \Phi^+ \Phi \theta_{i+1} 
\end{align}
We define the update vector $p_i$ by $\theta_{i+1} = \theta_i + \alpha p_i$. With this definition one get the reverse function definition:
\begin{align}
    \alpha B* p_i - p_i = I_i^{res},
\end{align}
where we have used the definition of the beam, and $I_i^{res} = \Phi^+ V - \Phi^+ \Phi \theta_i$ being the residual after the $i$th iteration. This equation again describes a deconvolution problem that needs to be performed to find the update vector $p_i$, similar as for traditional CLEAN, but this time with an updated convolution kernel $\alpha B - \delta$. We solve this deconvolution problem through a minor loop. This leads to the algorithm shown in Tab. \ref{tab: implicit_gd}. Comparing classical CLEAN with implicit gradient descent, we end up with a similar algorithm, just the beam used during the minor loop has been changed. Interestingly the addition of the delta function pushes the singular values of this new beam away from zero, making the deconvolution well-posed.

\begin{algorithm}
\caption{Implicit Gradient Descent CLEAN}

\begin{algorithmic}
\State $\theta_0 = 0$, $I_0^{res} = \Phi^+V$, $B = \Phi^+\Phi \delta$
\State $p_0 = \mathrm{minorloop}(I_0^{res}, B)$
\While{$k \mapsto k+1$ until convergence}
\State $\theta_{k+1} = \theta_k+\alpha p_k$
\State Update residual $I_{k+1}^{res}$ with new guess solution $\theta_{k+1}$
\State $p_{k+1} = \mathrm{minorloop}(I_{k+1}^{res}, \alpha B-\delta)$
\EndWhile
\end{algorithmic}

\label{tab: implicit_gd}
\end{algorithm}

\subsection{Conjugate Gradient Descent}
Now we follow the same logic as we did in Sec. \ref{sec: implicit_gd}, but apply it to a more sophisticated framework. In this subsection, we discuss the conjugate gradient descent method \citep[CG, e.g.][]{Nocedal2006}. CG solves a problem of the form $B\theta = b$, with a general, invertible matrix $B$ and some vectors $\theta$ and $b$. To this end, a sequence of vectors $v_1, ..., v_n$ is constructed which are orthogonal with respect to $B$, i.e. $v_i^+ B v_j = 0$ whenever $i \neq j$. These vectors define an orthogonal basis, and we can express the solution to our matrix inversion problem in this basis:
\begin{align}
    \theta = \sum_i \alpha_i v_i.
\end{align}
It follows:
\begin{align}
    v_k^+ b = v_k^+ B \theta = \sum_i \alpha_i v_k^+ B v_i = \alpha_k v_k^+ B v_k,
\end{align}
and hence:
\begin{align}
    \alpha_k = \frac{v_k^+ b}{v_k^+ B v_k}.
\end{align}
To solve a problem like $B\theta=b$ in an iterative form, CG creates such a sequence of orthonormal basis vectors through the following iterative steps:
\begin{enumerate}
    \item Choose as a first basis vector the initial residual: $p_0 = b - B \theta_0$.
    \item At an iteration $k$, compute the current residual: $r_k = b - B \theta_k$.
    \item Now find a vector orthonormal to the sequence of vectors $p_1, p_2, ..., p_{k-1}$. This can be done by classical Gram-Schmidt orthonormalization: $p_k = r_k - \sum_{i < k} \frac{p_i^+ B r_k}{p_i^+ B p_i} p_i$.
    \item Finally, update the model: $\theta_{k+1} = \theta_k + \alpha_k p_k$ with a scalar stepsize $\alpha_k > 0$ defined by: $\alpha_k = \frac{p_k^+ r_k}{p_k^+ B p_k}$. Loop over step 2.-4. until convergence is achieved.
\end{enumerate}

There are many different variants of the CG algorithm. For a more complete overview, as well as details on the method, its variants and convergence rate, we refer the reader to the textbook \citet{Nocedal2006}. Here we use the Fletcher-Reeves version \citep{Fletcher1964}. We note that one does not need to store all previous update directions. Rather, the orthogonalization could be built iteratively from the previous update direction by the update $p_{k+1} = r_{k+1} + \beta_k p_k$, where $\beta_k = \frac{r_{k+1}^+ r_{k+1}}{r_k^+ r_k}$ \citep{Nocedal2006}.

Numerical performance can be improved by the use of a preconditioner $M$ for the matrix $B$, i.e. a suitable matrix ensuring that $M^{-1} B$ has a smaller condition number than $B$. The general algorithm to solve the preconditioned optimization problem with the CG method is shown in Alg. \ref{tab: cgoptim}.

It is a common practice in radio-astronomy to introduce an additional visibility weighting. This weighting corrects for the bias introduced by a usually core-dominated sampling, over-emphasizing large scales in consequence \citep{Briggs1995}. These weighting schemes could in principle be absorbed into off-diagonal elements of the Gram matrix $G_{\mathbf{Y}}$, and thus in the operators $\Phi$ and $\Phi^+$. Hence, they affect the nature of the preconditioner. We refer the interested reader to the detailed, recent discussion in \citet{Bester2025} regarding the association of the preconditioner and the weighting scheme. \citet{Bester2025} shows that the discrete Fourier transform could be understood as an approximate eigenvalue decomposition of the Hessian with eigenvalues corresponding to Fourier transform of the point spread function. While this exact preconditioning may not be useful directly \citep[see for a detailed discussion][]{Bester2025}, it is observed that uniform weighting accelerates convergence by reducing the condition number of $M$ due to dividing out the largest eigenvalues.

We like to highlight two substeps in the algorithm. We solve the simpler system of linear equations $M z_k = r_k$ in every iteration. Moreover, we update the right hand side of the system of linear equation $r_k$ in every iteration by applying $B$ to the update vector of the guess solution, i.e. $r_{k+1} = r_k - \alpha_k Bp_k$. It is simple to show that this iteration yields $r_{k} = b - B \theta_{k}$, i.e. $r_k$ can be interpreted as the residual at iteration $k$.

Now we adapt the same idea as outlined in Sec. \ref{sec: implicit_gd}. We apply the general outline of the CG algorithm to the deconvolution problem $B \theta = I^{res}$, with the rank-deficit point spread function $B$. To translate these iterations into the CLEAN language, we simply solve the inversion of $M$ by a minor loop, and the update of the right hand side (the residual in the language of CLEAN) by the major loop iteration. The resulting algorithm is shown in Alg. \ref{tab: cg}.

The present motivation of the algorithm is not a formal derivation, hindered by the challenges to formulate CLEAN as a rigorous optimization method with a well-defined objective functional in general. The algorithm is however a natural and straightforward transfer of the ideas behind optimization through conjugate gradients into the CLEAN framework. We show in Appendix \ref{app: orthogonality} that the algorithm satisfies orthogonality of the search directions (at least approximately), i.e. effectively implements the idea of conjugate gradients.

\begin{algorithm}
\caption{CG Algorithm to solve linear equation system $B \theta = b$ with preconditioner $M$ for matrix $B$}

\begin{algorithmic}
\State $\theta_0 = 0$, $r_0 = b$
\State Solve $M z_0 = r_0$, $p_0 = z_0$
\While{$k \mapsto k+1$ until convergence}
\State $\alpha_k = \frac{r_k^+  z_k}{p_k^+ B p_k}$
\State $\theta_{k+1} = \theta_k+\alpha_k p_k$
\State $r_{k+1} = r_k - \alpha_k B p_k$
\State Solve $M z_{k+1} = r_{k+1}$
\State $\beta_k = \frac{r_{k+1}^{+} z_{k+1}}{r_{k}^{+} z_k}$, $p_{k+1} = z_{k+1}+\beta_k p_k$
\EndWhile
\end{algorithmic}

\label{tab: cgoptim}

\end{algorithm}

%\begin{table}
%\caption{CG CLEAN}

%\begin{tabular}{p{0.45\textwidth}}
%\hline \\
%\end{tabular}

%\begin{algorithmic}
%\State $\theta_0 = 0$, $I_0^{res} = \Phi^+V$, $B = \Phi^+\Phi \delta$
%\State $z_0 = \mathrm{minorloop}(I_0^{res}, B)$, $p_0 = z_0$
%\While{$k \mapsto k+1$ until convergence}
%\State $\alpha_k = \frac{I_k^{res,+}  z_k}{p_k^+ B p_k}$
%\State $\theta_{k+1} = \theta_k+\alpha_k p_k$
%\State Update residual $I_{k+1}^{res}$ with new guess solution $\theta_{k+1}$
%\State $z_{k+1} = \mathrm{minorloop}(I_{k+1}^{res}, B)$
%\State $\beta_k = \frac{I_{k+1}^{res,+} z_{k+1}}{I_{k}^{res,+} z_k}$, $p_{k+1} = z_{k+1}+\beta_k p_k$
%\EndWhile
%\end{algorithmic}

%\begin{tabular}{p{0.45\textwidth}}
%\hline \\
%\end{tabular}

%\label{tab: cg}

%\end{table}

\begin{algorithm}
\caption{CG CLEAN}

\begin{algorithmic}
\State $\theta_0 = 0$, $I_0^{res} = \Phi^+V$, $B = \Phi^+\Phi \delta$
\State $z_0 = \mathrm{minorloop}(I_0^{res}, B)$, $p_0 = z_0$
\While{$k \mapsto k+1$ until convergence}
\State $\alpha_k = \frac{I_k^{res,+}  p_k}{p_k^+ B p_k}$
\State $\theta_{k+1} = \theta_k+\alpha_k p_k$
\State Update residual $I_{k+1}^{res}$ with new guess solution $\theta_{k+1}$
\State $z_{k+1} = \mathrm{minorloop}(I_{k+1}^{res}, B)$
\State $\beta_k = -\frac{z_{k+1}^{+} B p_k}{p_k^+ B p_k}$, $p_{k+1} = z_{k+1}+\beta_k p_k$
\EndWhile
\end{algorithmic}

\label{tab: cg}

\end{algorithm}

\subsection{Momentum Acceleration}
A common way to accelerate gradient based optimization algorithms is by adding a momentum (also called the Nesterov-trick), resulting in algorithms such as the heavy-ball algorithm. In fact, the widely used Fast Iterative Soft Thrinkage Thresholding Algorithm (FISTA) is derived from momentum acceleration of classical forward-backward splitting with a $l^1$ regularization term. The idea, originating from a correspondence to a simple physics problem, is keeping a fraction of the momentum of the previous gradient update for the next update. And indeed, such an approach seems reasonable for radio-interferometry in any situation where the residual decreases from iteration to iteration, but keeps a similar structure, i.e. the gradient approximately keeps its direction.

The Newton setting that we (albeit inaccurately) used to understand CLEAN, is derived from a Taylor expansion among the current solution. We can introduce a momentum by evaluating the Taylor sum at the location $\theta_{k}+\mu v_k$ rather than $\theta_k$. Here $\mu \in [0,1]$ is a real number smaller than one, and $v_k$ describes the last update step, i.e. $\theta_k = \theta_{k-1}+v_k$. We get the Taylor expansion to second order for a model solution $M$:
\begin{align} \nonumber
     J(\hat{\theta}) \approx &J(\theta_k+\mu v_k) + \nabla J\left( \hat{\theta}-\theta_k-\mu v_k\right) \\
     &+ \frac{1}{2} \left( \hat{\theta}-\theta_k-\mu v_k\right)^T \nabla^2 J \left( \hat{\theta}-\theta_k-\mu v_k\right).
\end{align}
Searching for the zeros of the gradient, we find the iteration:
\begin{align}
    \theta_{k+1} = \theta_{k} + \mu v_k +\left( \nabla^2 J \right)^{-1} \nabla J\left( \theta_k+\mu v_k\right).
\end{align}
This policy is implemented by algorithm \ref{tab: momentum_newton}. For better comparability to the CLEAN version of this algorithm, we have splitted the inversion of the Hessian and the calculation of the gradient here in two separated steps.

Applying the same arguments as in Sec. \ref{sec: CLEAN_Newton}, just backwards, we can design the momentum accelerated CLEAN version. We replace the calculation of the gradient with the major loop, and the inversion of the Hessian (i.e. the beam) by the minor loop). We show the pseudocode in Alg. \ref{tab: momentum}. The resulting algorithm is very similar to traditional CLEAN, with two small differences. We are evaluating the major loop iteration not at the current model, but at the model $\theta_k+\mu v_k$, and the model update is amended with a fraction of the previous model update step. It is directly visible from the pseudocode that none of the additional steps affects the computational complexity of neither the minor loop nor the residual-update step: The additional step simply involves summing two images, a step with the numerical complexity of a single minor loop iteration.

\begin{algorithm}
\caption{Momentum accelerated Newton Minimization of an objective $J$.}

\begin{algorithmic}
\State $\theta_0 = 0$, $I_0^{res} = \nabla J (\theta_0)$
\State $v_0 = 0$
\State $p_0 = (\nabla^2 J)^{-1} I_0^{res}$
\While{$k \mapsto k+1$ until convergence}
\State $v_{k+1} = \mu v_k + p_k$
\State $\theta_{k+1} = \theta_k+v_{k+1}$
\State Update residual $I_{k+1}^{res} = \nabla J (\theta_{k+1})$
\State $p_{k+1} = (\nabla^2 J)^{-1} I_{k+1}^{res}$
\EndWhile
\end{algorithmic}

\label{tab: momentum_newton}
\end{algorithm}

\begin{algorithm}
\caption{Momentum CLEAN}

\begin{algorithmic}
\State $\theta_0 = 0$, $I_0^{res} = \Phi^+V$, $B = \Phi^+\Phi \delta$
\State $v_0 = 0$
\State $p_0 = \mathrm{minorloop}(I_0^{res}, B)$
\While{$k \mapsto k+1$ until convergence}
\State $v_{k+1} = \mu v_k + p_k$
\State $\theta_{k+1} = \theta_k+v_{k+1}$
\State Update residual $I_{k+1}^{res}$ with guess model $\theta_{k+1}+\mu v_{k+1}$
\State $p_{k+1} = \mathrm{minorloop}(I_{k+1}^{res}, B)$
\EndWhile
\end{algorithmic}

\label{tab: momentum}
\end{algorithm}

\section{Validation}\label{sec:validation}

We test the different algorithms with a variety of observational data sets. We validate the algorithms hierarchically in four different steps of a validation ladder. These are:
\begin{itemize}
    \item First, we apply our algorithms to wideband, continuum observations with the VLA at C-band. These observations were taken during early science shared risk observing mode and have been prepared by the CASA team especially for the CASA continuum imaging tutorial\footnote{\url{https://casaguides.nrao.edu/index.php/VLA_Continuum_Tutorial_3C391-CASA6.4.1}}. Due to CASA's widespread usage in the radio-interferometry community and frequently occurring calibration and imaging workshops based on these tutorials, this dataset has evolved into one of most often utilized standard datasets for the community to test and commission new algorithms. For more details, we refer to aforementioned tutorial. We would like to mention however, that the dataset represents a real dataset, with identified issues at multiple antennas, and radio frequency inference (RFI) artifacts becoming apparent at large dynamic range. We followed the calibration steps outlined in the tutorial, and replaced the imaging part with our new algorithms Momentum-CLEAN and CG-CLEAN.
    \item The second step in our data validation consists of synthetic data with known ground truth. We used a real observed and reconstructed image of Cygnus A \footnote{selfcalibrated data shared in private communication by Rick Perley, Program Code: 14B-336} and a snapshot of the ENZO simulation \citep{Gheller2022} as ground truth images. The ENZO simulated galaxy image has been scaled up in size to match the spatial scales accessible to the VLA. The ground truth images are observed synthetically at 2 GHz with the VLA in A configuration in a narrow bandwidth of 128 MHz. We added atmospheric and thermal noise according to the system temperature measurements usual for the VLA in this frequency setup and elevation. The synthetic datasets have been recovered with various algorithms, and compared against the ground truth images.
    \item Next we turn our attention to narrow-band real datasets. To aid comparability, we used data sets of well-known sources that are traditionally used to benchmark algorithmic performances \citep[see e.g.][]{Arras2021, Dabbech2021, Hsieh2021, Roth2023, Dabbech2024, Mueller2025}. These are radio observations of the Andromeda galaxy M31, of the supernova remnant G055.7+3.4 and the radio galaxy Cygnus A with the VLA. The datasets are the same that were used in \citet{Hsieh2021}. We refer to this manuscript for further details on the observations and calibration. These datasets are nowhere near to perfectly calibrated datasets, but represent what could be expected from observed datasets with real data corruptions.
    \item Finally, we test the extendibility of the algorithms towards more challenging domains. For this purpose, we apply our algorithms in a high resource setting to a wideband mosaic of the galaxy NGC 628. The source has been observed with the VLA in L-band in semester 2023 B in D configuration \footnote{Program code: VLA-23B-025, PI: Karin Sandstrom}. For more details on the obervation, we refer to \citet{Sandstrom2023a, Sandstrom2023b}. Vice versa, we are also applying the algorithms to the low resource setting, i.e. where the number of visibilities and signal-to-noise ratio drops. To this end, we have used a randomly selected epoch\footnote{19th August 2018} from the MOJAVE monitoring of the blazar 3C 120 at 15 GHz with the VLBA \citep{Lister2018}. MOJAVE is one of the largest programs regularly observed with the VLBA. It utilizes the VLBA for snapshot observations of a set of AGN sources to track their motion over time. 3C 120 is one of the famous sources that exhibits a long jet and significant motion in between the different epochs \citep{Lister2018}, and in consequence is among the most often studied AGN sources in VLBI. We have selected this source explicitly because it has been monitored continuously for almost three decades by MOJAVE and its predecessors, which allows for comparison of our individual epoch reconstructions, to a stacked image from 30 years of monitoring and over 100 epochs \citep{Lister2018}.
\end{itemize}

We cleaned the datasets with three different algorithm architectures: The classical Cotton-Schwab CLEAN, CG-CLEAN (see Alg. \ref{tab: cg}), and Momentum-CLEAN (see Alg. \ref{tab: momentum}), both presented in this manuscript. For Momentum-CLEAN, we applied a momentum transfer described by a factor $\mu = 0.5$. Both CG-CLEAN and Momentum-CLEAN have been implemented in CASA \citep{CASA2022}. The major loop and minor loop are implemented by calls to CASA's \textit{tclean} function. Any additional computation, such as the calculation of the coefficients $\alpha$ and $\beta$ for CG-CLEAN, was realized with CASA's \textit{immath} package at the top level. 

This manuscript primarily focuses on the convergence speed of the major loop. However, the total number of major loop iterations required to reach convergence is also strongly affected by the choice of algorithm used in the minor loop. For example, it is well established that multiscale algorithms which perform deeper cleaning during each minor loop, such as Asp-CLEAN \citep{Bhatnagar2004, Hsieh2021}, MS-CLEAN \citep{Cornwell2008, Rau2011, Mueller2023a} and the recently proposed Autocorr-CLEAN \citep{Mueller2025}, can significantly reduce the number of major loop iterations required. Among the various multiscale imaging algorithms proposed as alternatives to the traditional CLEAN minor loop, we employ the best-performing methods currently available in the official CASA releases, specifically, MS-CLEAN and Asp-CLEAN, which have been identified by the CASA committee as the most robust options.

We like to highlight that it is part of the ongoing success and popularity of the CLEAN algorithm within the radio-interferometry community that robust, and proven metaheuristics in the form of automasking, adjusting the CLEAN gain, self-calibration, scale biases, primary beam limits and stopping rules fine-tuned to the multiscale scenario have been derived and validated to yield correct results in most scenarios, even for badly calibrated data. We have chosen the hyper-parameters controlling MS-CLEAN and Asp-CLEAN to the best of our knowledge, and use the same configuration for MS-CLEAN, Momentum-CLEAN and CG-CLEAN for every dataset to allow for same-level comparisons. For the datasets in the third step of our validations, we used configurations consistent with the configuration used in \citet{Hsieh2021}. Particularly for Asp-CLEAN these choices represent significant manual efforts to fine-tune the algorithm to the dataset at hand. Note that both MS-CLEAN and Asp-CLEAN have heuristical stopping rules implemented in \textit{CASA}, stopping the minor loop whenever it is diverging, frozen or adds large scale negative flux. For this analysis, we rely on these stopping criteria, i.e. in every minor loop we run MS-CLEAN and Asp-CLEAN as deep as possible without introducing divergence.

\section{Results}\label{sec:results}
In the following subsections, we present our results for every stage of our four stage validation ladder. Finally, we discuss hyper-parameter choices for Momentum-CLEAN.

\subsection{CASA tutorial}
\begin{figure*}
    \centering
    \includegraphics[width=\textwidth]{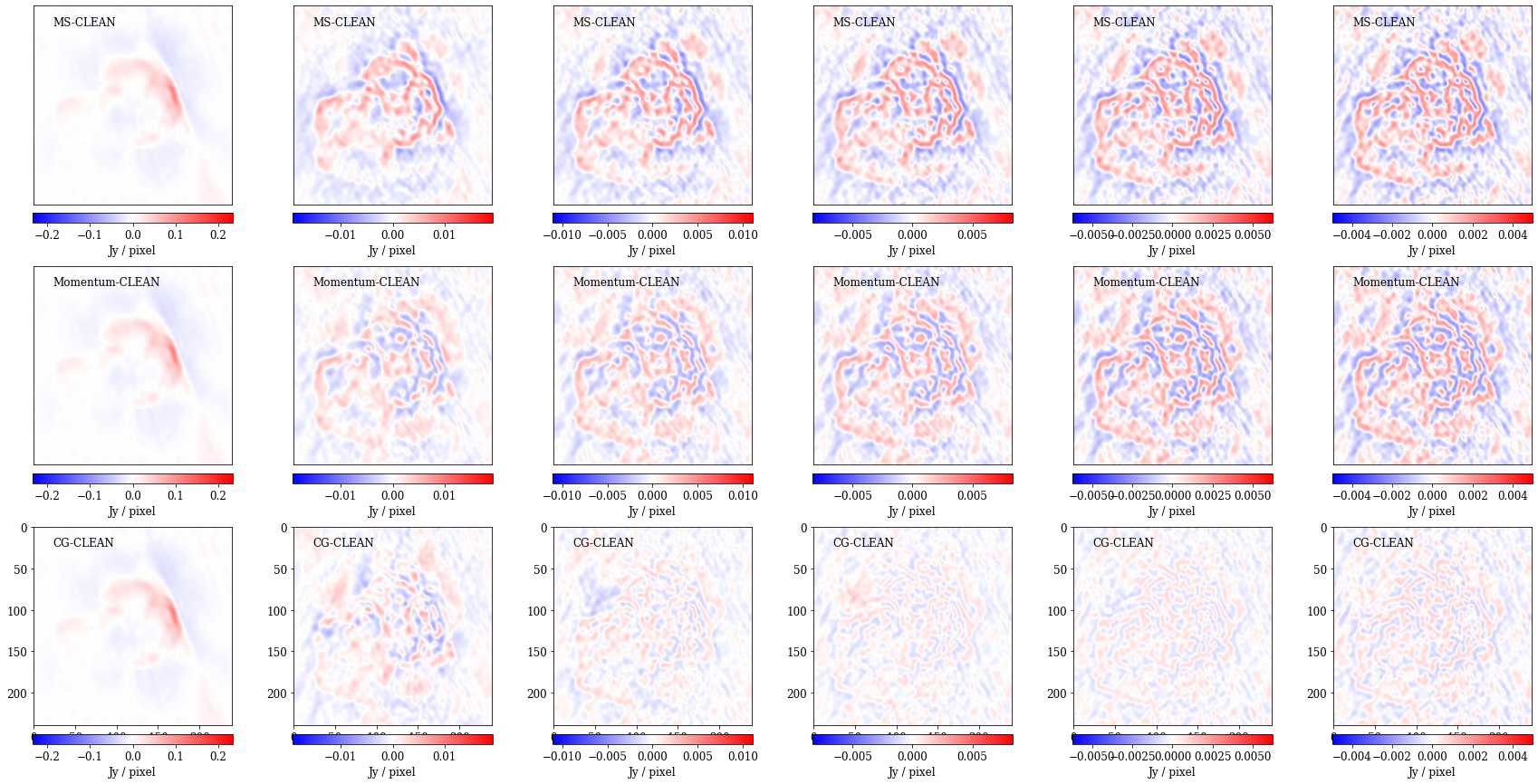}
    \caption{Residual as a function of increasing major loop iterations (from left to right) for 3C391 for CLEAN (top row), Momentum CLEAN (middle row) and CG-CLEAN (bottom row).}
    \label{fig:3C391}
\end{figure*}

\begin{figure*}
    \centering
    \includegraphics[width=\textwidth]{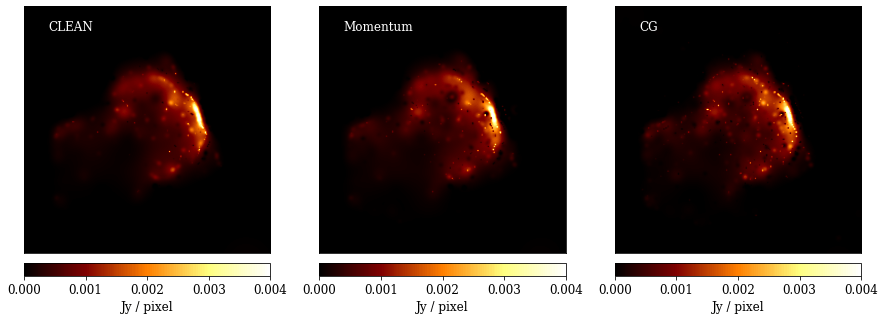}
    \caption{The recovered image of 3C391 with MS-CLEAN (left panel), Momentum-CLEAN (middle panel) and CG-CLEAN (right panel).}
    \label{fig:3C391_models}
\end{figure*}

\begin{figure*}
    \centering
    \includegraphics[width=\textwidth]{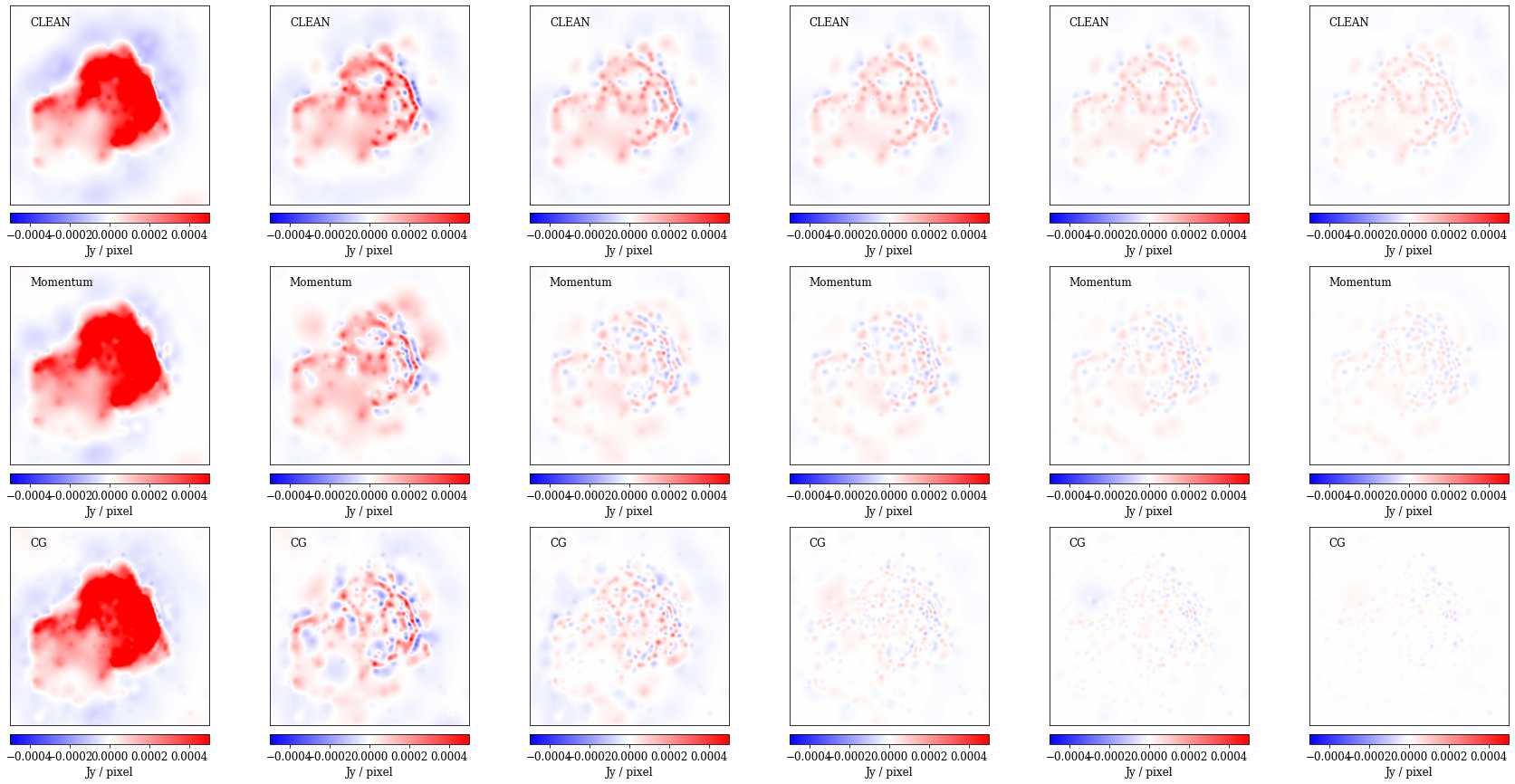}
    \caption{Difference between current model and final model (after the last major loop iteration) as a function of increasing major loop iterations (from left to right) for 3C391 for CLEAN (top row), Momentum CLEAN (middle row) and CG-CLEAN (bottom row).}
    \label{fig:3C391_models_iter}
\end{figure*}

We show the residual as a function of major loop iteration in Fig. \ref{fig:3C391}, and the final model in Fig. \ref{fig:3C391_models}. Finally, we present the difference between the current model and final model, i.e. $\theta_N - \theta_k$ (where $N$ is the number of major loops run in total) following the notation of Alg. \ref{tab: clean}, \ref{tab: cg} and \ref{tab: momentum}, as a function of major loop iteration (i.e. of $k$). Our claim for convergence of the algorithms stems from two observations:
\begin{itemize}
\item The residual which measures the fidelity between the model and the data converges faster to a noise-like distribution for CG-CLEAN than for classical CLEAN, and approximately with the same speed for Momentum-CLEAN. We would like to highlight that while the residual in principle does not evaluate the match between the reconstructed solution and true emission structure, it is typically used as an indicator for the success of the CLEANing procedure and as metric to compare performance of algorithms nevertheless \citep[e.g.][]{Offringa2017, Hsieh2021, Bester2025}. This is reasonable since due to the heuristic restrictions applied throughout the application of CLEAN (i.e. stopping rules and windows), it is usually disfavored to put emission in unphysical location or in a noise-like pattern overfitting the data.
\item The final models for the reconstructions for all three algorithms match well, as shown in Fig. \ref{fig:3C391_models}. While the validity of images obtained by CLEAN in general could be disputed as well, these images represent the de-facto standard, generally accepted to represent the truth in the community. By inspecting Fig. \ref{fig:3C391_models_iter}, we note however that we need less major loop iterations to achieve these consensus solutions, with CG-CLEAN performing best, followed by Momentum-CLEAN, and traditional CLEAN being the slowest algorithm.
\end{itemize}
In summary, the algorithms presented in this manuscript produce smaller residuals in fewer iterations, arrive at the same solution as traditional CLEAN, but need fewer major loop iterations for that, i.e. lead to significant acceleration.

\subsection{Synthetic data}
\begin{figure*}
    \centering
    \includegraphics[width=\textwidth]{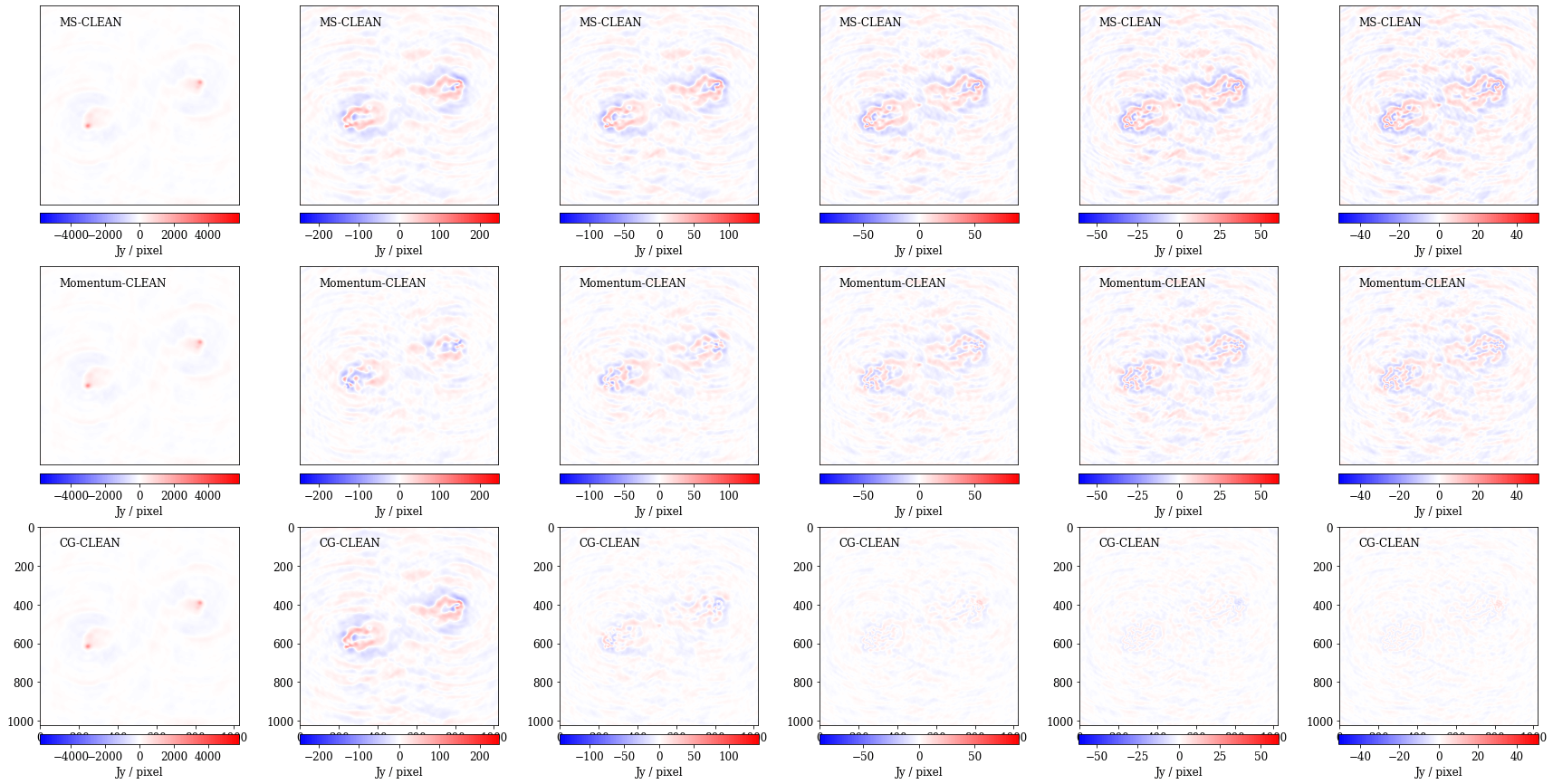}
    \caption{Residual as a function of increasing major loop iterations (from left to right) for synthetic Cygnus A data for CLEAN (top row), Momentum CLEAN (middle row) and CG-CLEAN (bottom row).}
    \label{fig:synth_cygnus}
\end{figure*}

\begin{figure*}
    \centering
    \includegraphics[width=\textwidth]{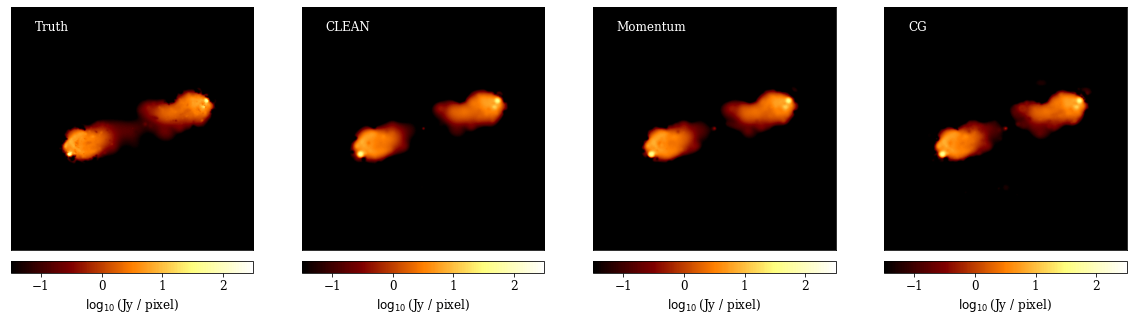}
    \caption{The ground truth image (left panel) and the recovered images of synthetic Cygnus A data with MS-CLEAN (second panel), Momentum-CLEAN (third panel) and CG-CLEAN (right panel).}
    \label{fig:synth_cygnus_models}
\end{figure*}

\begin{figure*}
    \centering
    \includegraphics[width=\textwidth]{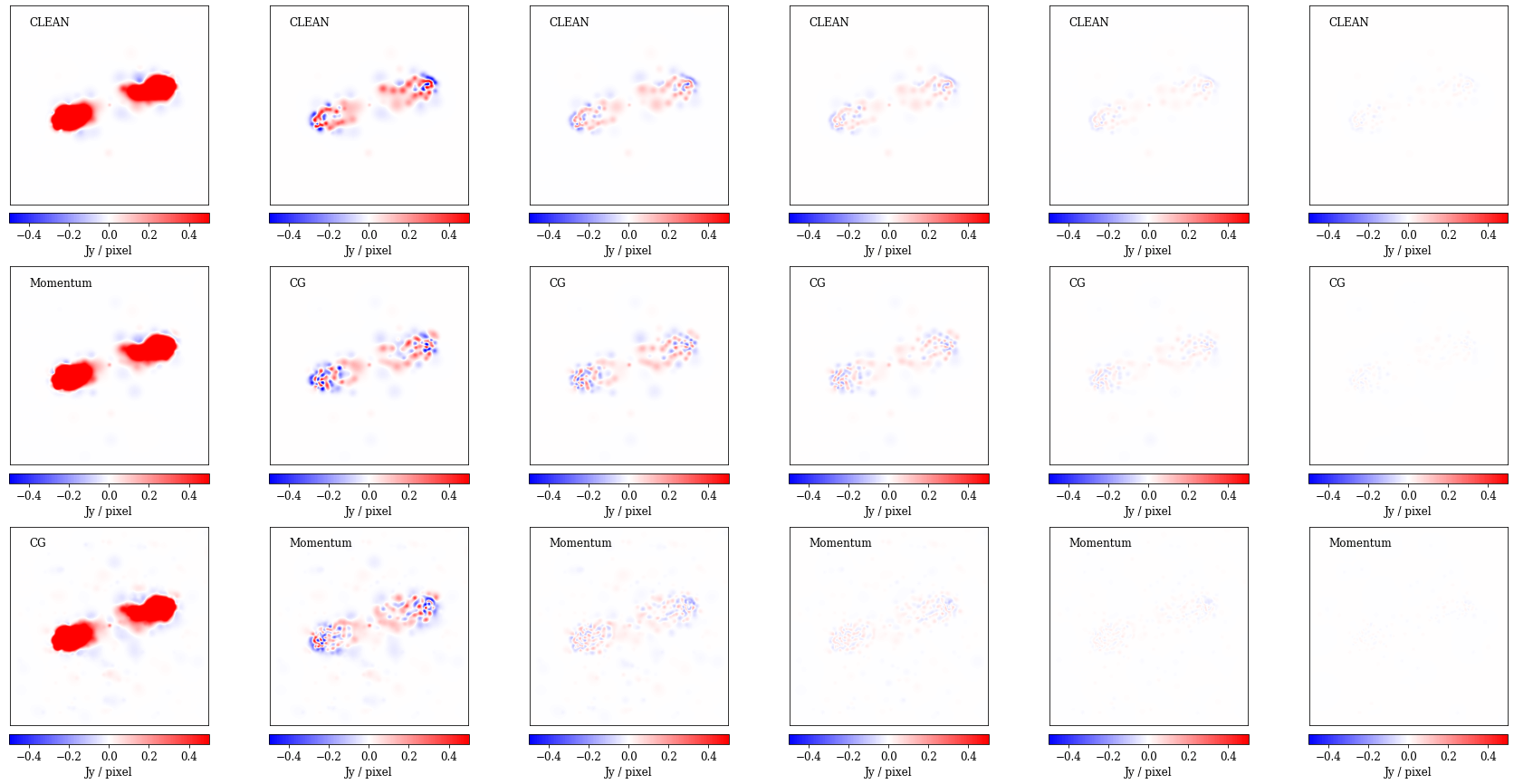}
    \caption{Difference between current model and final model (after the last major loop iteration) as a function of increasing major loop iterations (from left to right) for synthetic ENZO simulation data for CLEAN (top row), Momentum CLEAN (middle row) and CG-CLEAN (bottom row).}
    \label{fig:synth_cygnus_models_iter}
\end{figure*}

\begin{figure*}
    \centering
    \includegraphics[width=\textwidth]{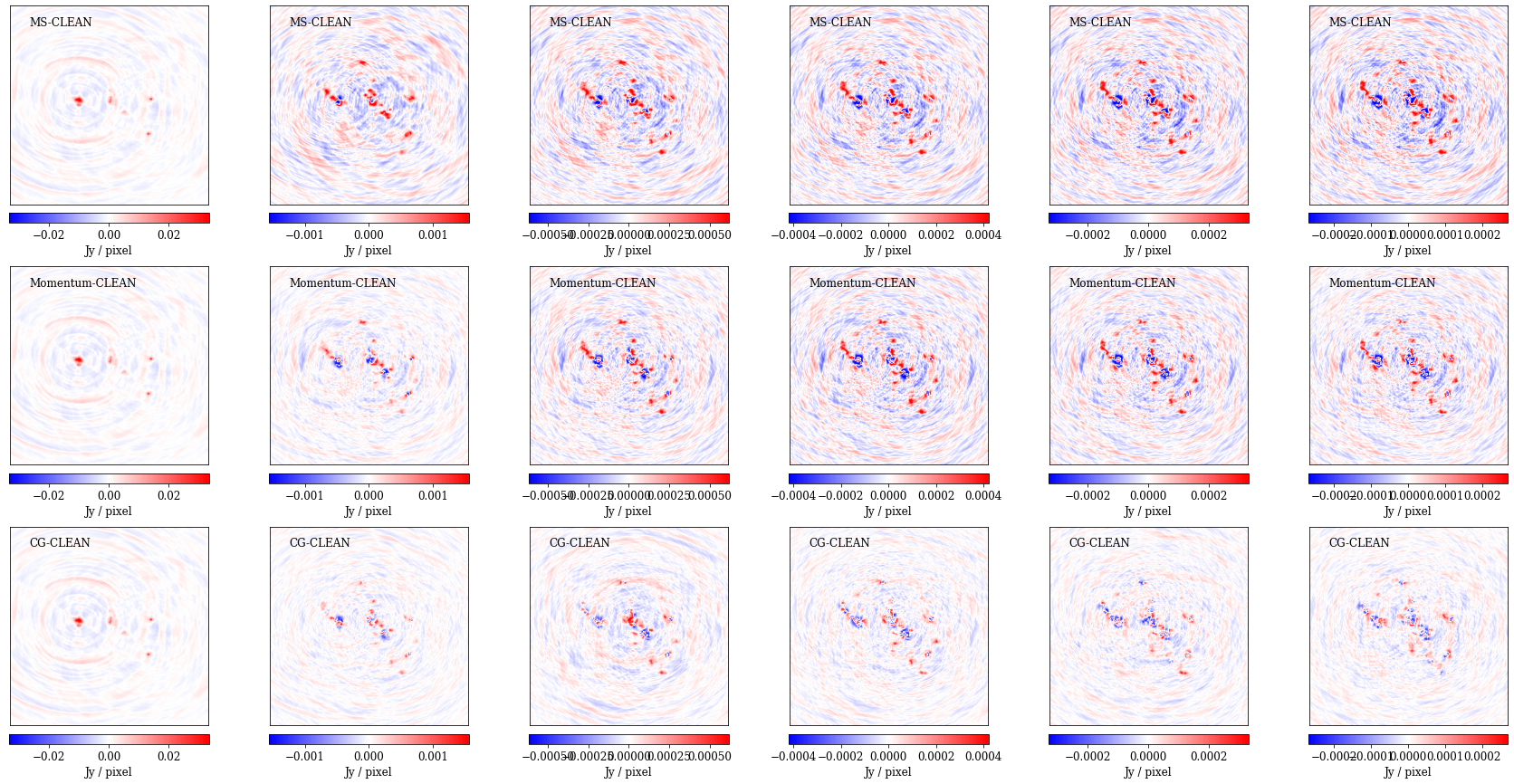}
    \caption{Residual as a function of increasing major loop iterations (from left to right) for synthetic ENZO simulation data for CLEAN (top row), Momentum CLEAN (middle row) and CG-CLEAN (bottom row).}
    \label{fig:synth_enzo}
\end{figure*}

\begin{figure*}
    \centering
    \includegraphics[width=\textwidth]{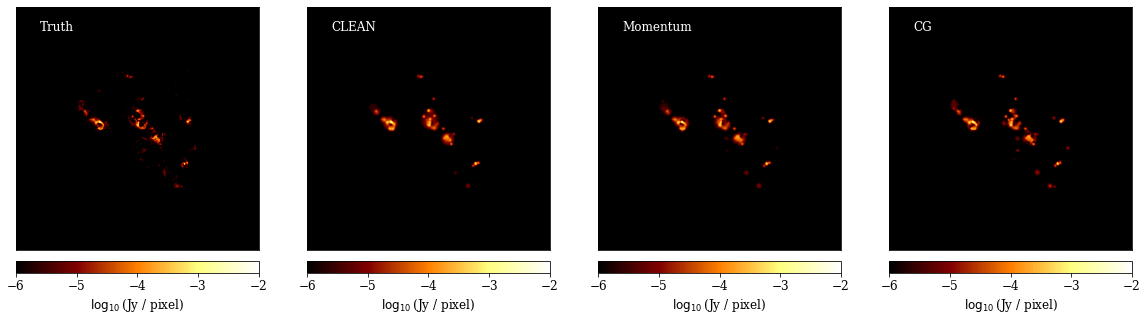}
    \caption{The ground truth image (left panel) and the recovered images of synthetic ENZO simulation data with MS-CLEAN (second panel), Momentum-CLEAN (third panel) and CG-CLEAN (right panel).}
    \label{fig:synth_enzo_models}
\end{figure*}

\begin{figure*}
    \centering
    \includegraphics[width=\textwidth]{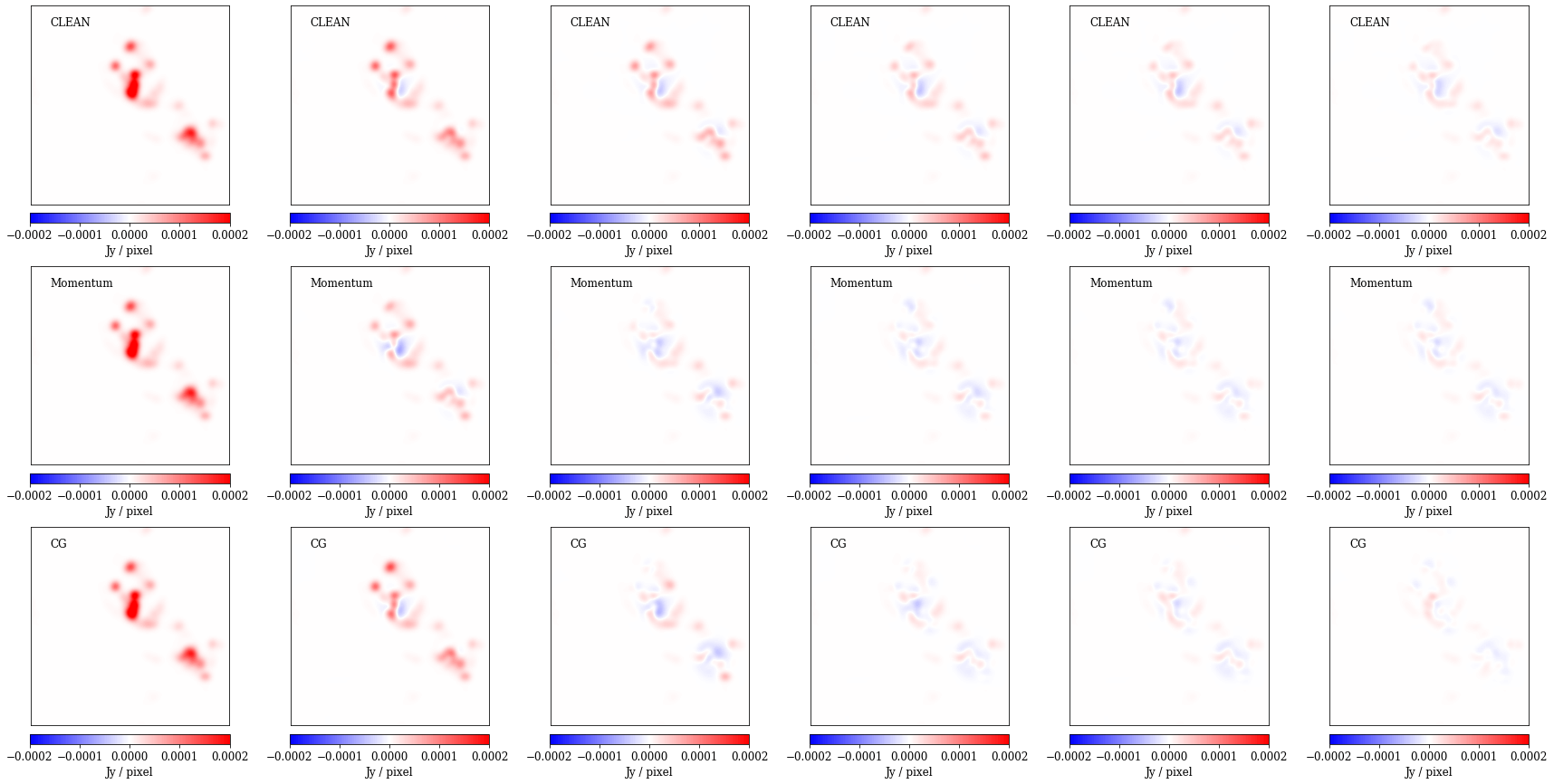}
    \caption{Difference between current model and final model (after the last major loop iteration) as a function of increasing major loop iterations (from left to right) for synthetic Cygnus A data for CLEAN (top row), Momentum CLEAN (middle row) and CG-CLEAN (bottom row).}
    \label{fig:synth_enzo_models_iter}
\end{figure*}

As in previous subsection, we present the residuals depending on the major loop iteration, the final model and the difference between the final model and current model varying with the number of major loop iterations for the synthetic data of Cygnus A in Fig. \ref{fig:synth_cygnus}, \ref{fig:synth_cygnus_models} and Fig. \ref{fig:synth_cygnus_models_iter}, and for the simulated galaxy image from the ENZO simulation in \ref{fig:synth_enzo}, \ref{fig:synth_enzo_models} and Fig. \ref{fig:synth_enzo_models_iter}. We observe the same findings that we have reported in the previous subsection. CG-CLEAN and Momentum-CLEAN converge towards results that are compatible to classical CLEAN, without obvious artifacts. However, for CG-CLEAN we need less major loop iterations to achieve this solution, as well as less iterations to get a smaller residual. Momentum-CLEAN offers marginal advantages over traditional CLEAN for these test examples by visual examination.

Synthetic data sets with known ground truth offer another option to substantiate the claim of acceleration of the traditional CLEAN algorithm: We can directly compare the reconstructions to the ground truth images. First, by visibly examining Fig. \ref{fig:synth_cygnus_models} and Fig. \ref{fig:synth_enzo_models}, we do not see obvious artifacts caused by the CG-CLEAN and Momentum-CLEAN procedures. The recovered images represent the ground truth images reasonably well. One could already see from these comparisons that CG-CLEAN is actually providing a slightly better reconstruction, e.g. the inner wisps of the termination lobes in Cygnus A are better represented by CG-CLEAN than by classical Cotton-Schwab CLEAN.

\begin{figure*}[t!]
    \centering
    \begin{subfigure}[b]{0.45\textwidth}
        \centering
        \includegraphics[width=\textwidth]{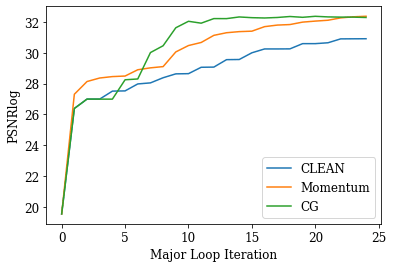}
        \caption{Cygnus A}
    \end{subfigure}
    \begin{subfigure}[b]{0.45\textwidth}
        \centering
        \includegraphics[width=\textwidth]{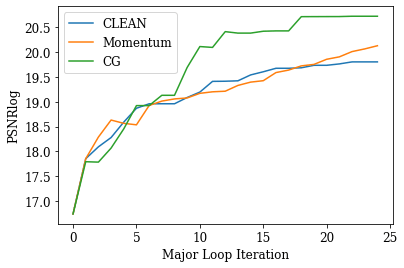}
        \caption{ENZO galaxy}
    \end{subfigure}
    \caption{PSNR of the reconstructions as a function of major loop iteration for synthetic data of Cygnus A (left panel) and a simulated galaxy from the ENZO simulation suite (right panel).}
    \label{fig:psnrs}
\end{figure*}

To quantify the accuracy of the algorithms we use the widely adopted PSNR metric. The metric is defined by:
\begin{align}
\mathrm{PSNR} = 20 \log_{10} \left( \frac{\mathrm{max}(I)}{||\theta-I||} \right).
\end{align}
Here $I$ is the ground truth image and $\theta$ the reconstruction. The PSNR metric however is dominated by the reconstruction fidelity of bright point sources, such as the termination shocks in Cygnus A. We therefore utilize an alternative which emphasizes the fidelity of the fainter, diffuse emission component, by evaluating the logarithms (the logarithms have been thresholded by the noise-level):
\begin{align}
\mathrm{PSNRlog} = 20 \log_{10} \left( \frac{\mathrm{max}(\log I)}{||\log \theta-\log I||} \right).
\end{align}
We show the accuracy of the recovered models as a function of major loop iteration in Fig. \ref{fig:psnrs}. For Cygnus A, CG-CLEAN and Momentum-CLEAN are the best performing algorithms regarding the precision of the final recovered model, outperforming CLEAN. In the ENZO example, CG-CLEAN outperforms both Momentum-CLEAN and CLEAN. In terms of speed, we note that Momentum-CLEAN leads to the fastest algorithm in the first few iterations. CG-CLEAN shows a similar speed as CLEAN in the first few iterations, but then overtakes both Momentum-CLEAN and CLEAN. A natural explanation could be that the orthogonalization of search directions inherent to CG-CLEAN starts to be an effective acceleration only after a few iterations when the Krylov subspace gets larger and more informative. Strikingly, we observe that with CG-CLEAN we achieve the accuracy of CLEAN in approximately a fifth of major loop iterations, backed by similar acceleration factors obtained from the size of the residual.

\subsection{Real narrowband observations}
Now we turn to the third stage of our validation ladder, containing real, narrow-band observations with the VLA. We selected to test sources at this stage that are frequently used as validation data sets. However, the observations studied in this subsection do not represent perfectly calibrated data sets, but real observational data sets with residual calibration errors.

\begin{figure*}
    \centering
    \includegraphics[width=\textwidth]{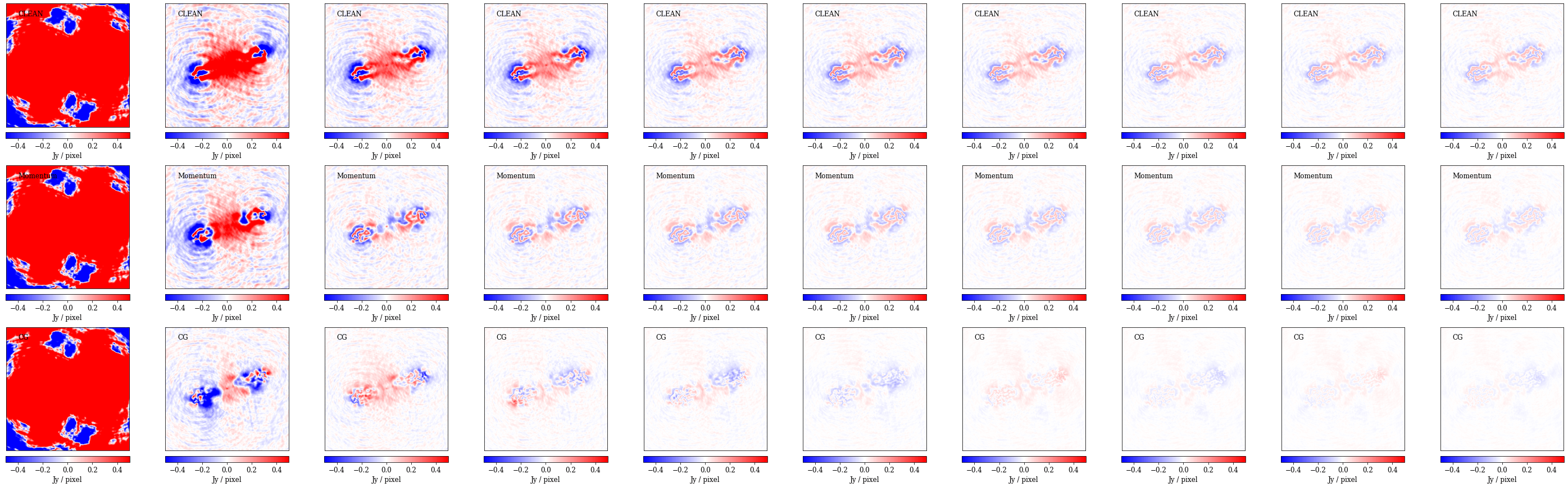}
    \caption{Residual as a function of increasing major loop iterations (from left to right) for Cygnus A for CLEAN (top row), Momentum CLEAN (middle row) and CG-CLEAN (bottom row).}
    \label{fig:cygA}
\end{figure*}

\begin{figure*}
    \centering
    \includegraphics[width=\textwidth]{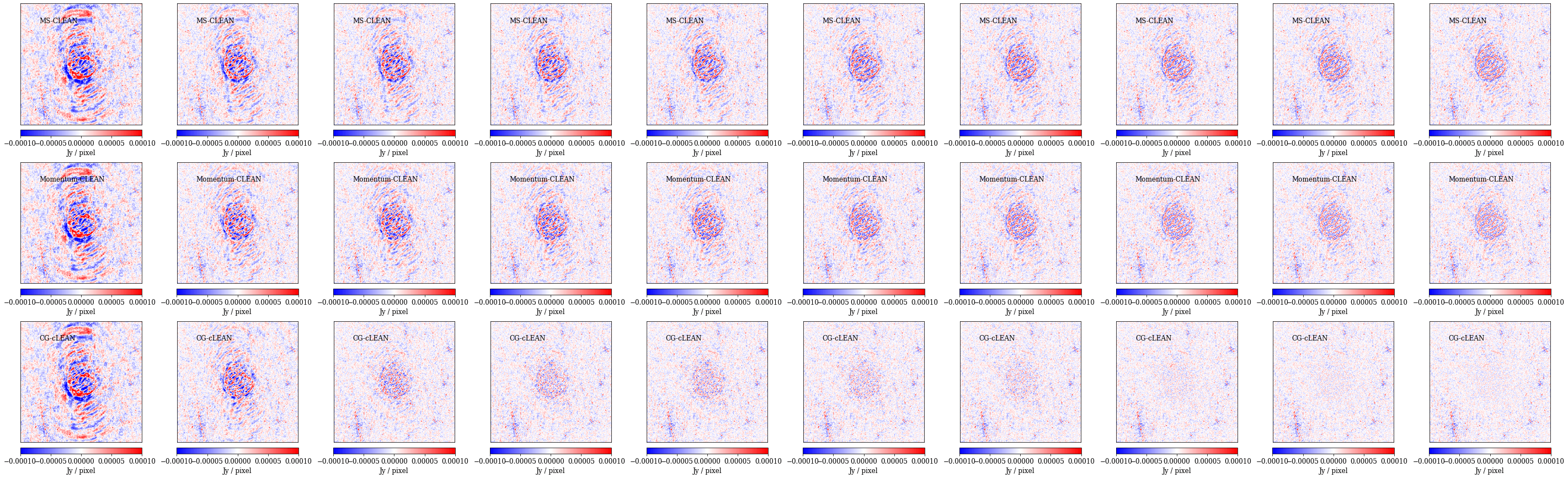}
    \caption{Same as Fig. \ref{fig:cygA}, but for G055.7+3.4.}
    \label{fig:g55}
\end{figure*}

\begin{figure*}
    \centering
    \includegraphics[width=\textwidth]{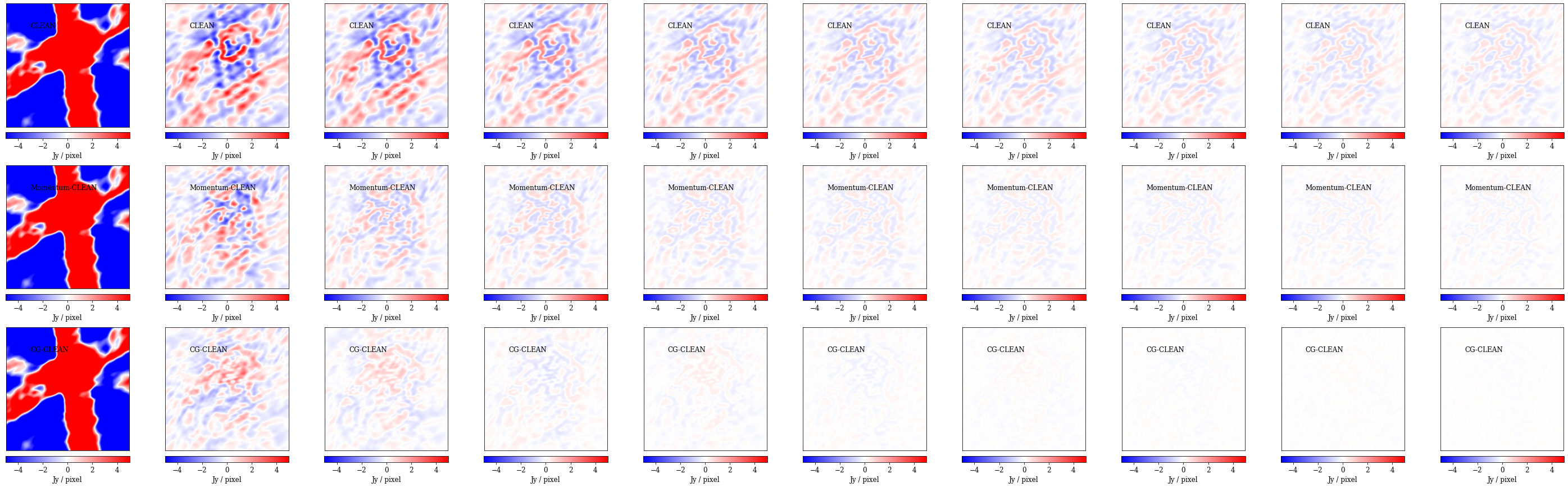}
    \caption{Same as Fig. \ref{fig:cygA}, but for M31.}
    \label{fig:m31}
\end{figure*}

\begin{figure*}[t!]
    \centering
    \begin{subfigure}[b]{\textwidth}
        \centering
        \includegraphics[width=\textwidth]{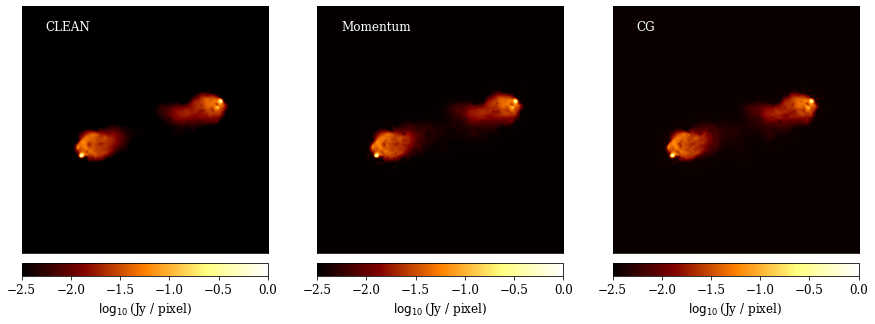}
        \caption{Cygnus A}
    \end{subfigure}%
    \\
    \begin{subfigure}[b]{\textwidth}
        \centering
        \includegraphics[width=\textwidth]{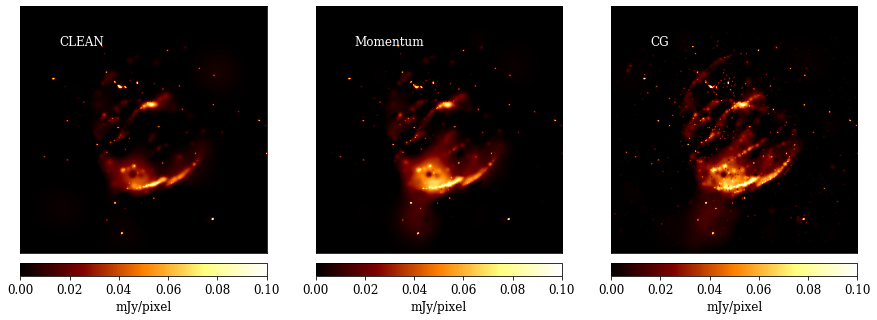}
        \caption{G055.7+3.4}
    \end{subfigure}
        \\
    \begin{subfigure}[b]{\textwidth}
        \centering
        \includegraphics[width=\textwidth]{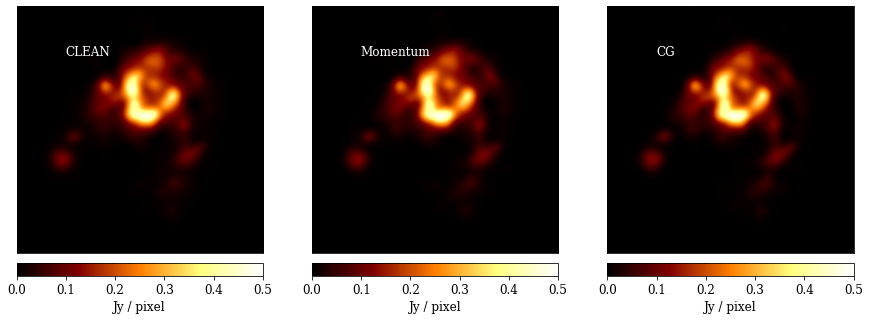}
        \caption{M31}
    \end{subfigure}
    \caption{Recovered models with CLEAN (left column), Momentum CLEAN (middle column), and CG-CLEAN (right column) for Cygnus A (first row), G055.7+3.4 (middle row), and M31 (bottom row).}
    \label{fig:model}
\end{figure*}

\begin{figure*}
    \centering
    \includegraphics[width=\textwidth]{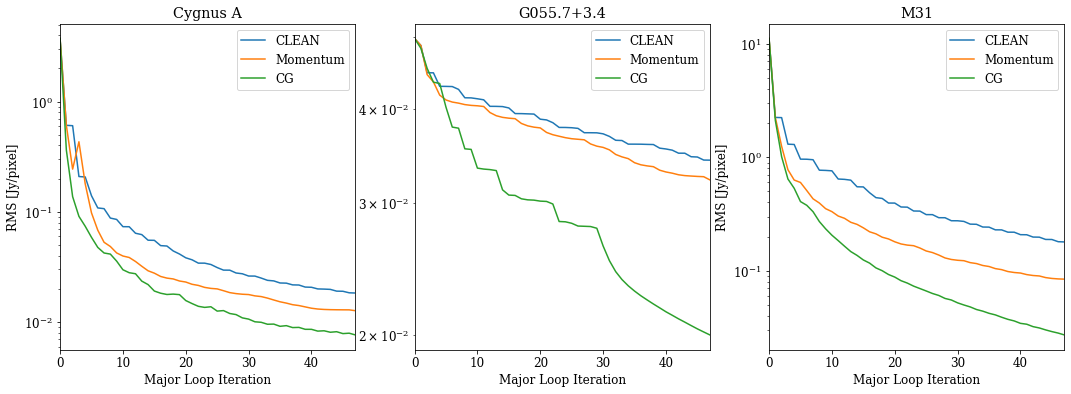}
    \caption{Norm of the residual as a function of major loop iteration for CLEAN (blue), Momentum CLEAN (orange) and CG-CLEAN (green). The panels show from left to right the results for Cygnus A, G055.7+3.4 and M31.}
    \label{fig:rms}
\end{figure*}

We show in Fig. \ref{fig:cygA}, \ref{fig:g55} and \ref{fig:m31} the residual for Classical CLEAN, Momentum-CLEAN and CG-CLEAN for Cygnus A, G055.7+3.4 and M31. The left column shows the initial, first residual. From left to right, we show the residual with progressing number of major loop iterations. Moreover, we present the recovered models in Fig. \ref{fig:model}, and present a more quantitative comparison of the norm of the residual as a function of major loop iteration in Fig. \ref{fig:rms}.

First of all, we recognize that all algorithms are successful in cleaning the structure. The residual continuously decreases without introducing artifacts. The structure seen for CLEAN to create residuals with negative bowls around bright pointy sources (e.g. visible from the termination shocks in Cygnus A), is well documented, and could potentially be counteracted by finetuning the scale bias. 

Momentum-CLEAN converges more rapidly than standard CLEAN across all three examples in terms of size of residual per major loop iteration, highlighting the substantial benefit of incorporating a fraction of the previous update step into each subsequent model update.

CG-CLEAN significantly outperforms both standard CLEAN and Momentum-CLEAN across all three test sources. This holds true in terms of both convergence speed and the overall depth reached by the algorithm. Specifically, CG-CLEAN achieves the same dynamic range as CLEAN with 3–5 times fewer major loop iterations and continues to attain higher dynamic ranges in later iterations. These findings related to the performance of CG-CLEAN and Momentum-CLEAN are consistent with the findings that we reported for the first two rungs of our validation ladder.

\begin{figure*}
    \centering
    \includegraphics[width=\textwidth]{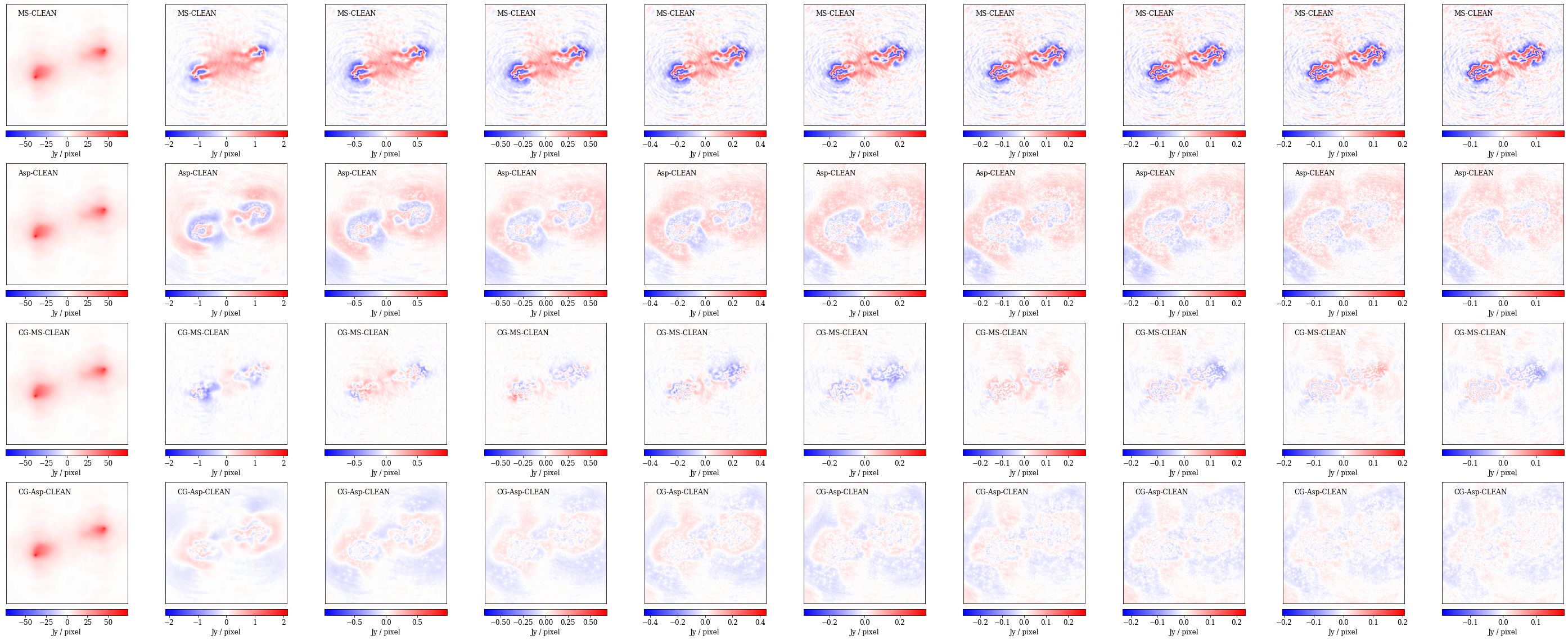}
    \caption{Residual as a function of major loop iteration for Cygnus A. Here, we compare traditional CLEAN with CG-CLEAN in two different flavors for the minor loop. First row: Cotton-Schwab major cycles with MS-CLEAN in the minor loop. Second row: Cotton-Schwab major cycles with Asp-CLEAN in the minor loop. Third row: CG-CLEAN with MS-CLEAN in the minor loop. Fourth row: CG-CLEAN with Asp-CLEAN in the minor loop.}
    \label{fig:cygA_asp}
\end{figure*}

\begin{figure*}
    \centering
    \includegraphics[width=\textwidth]{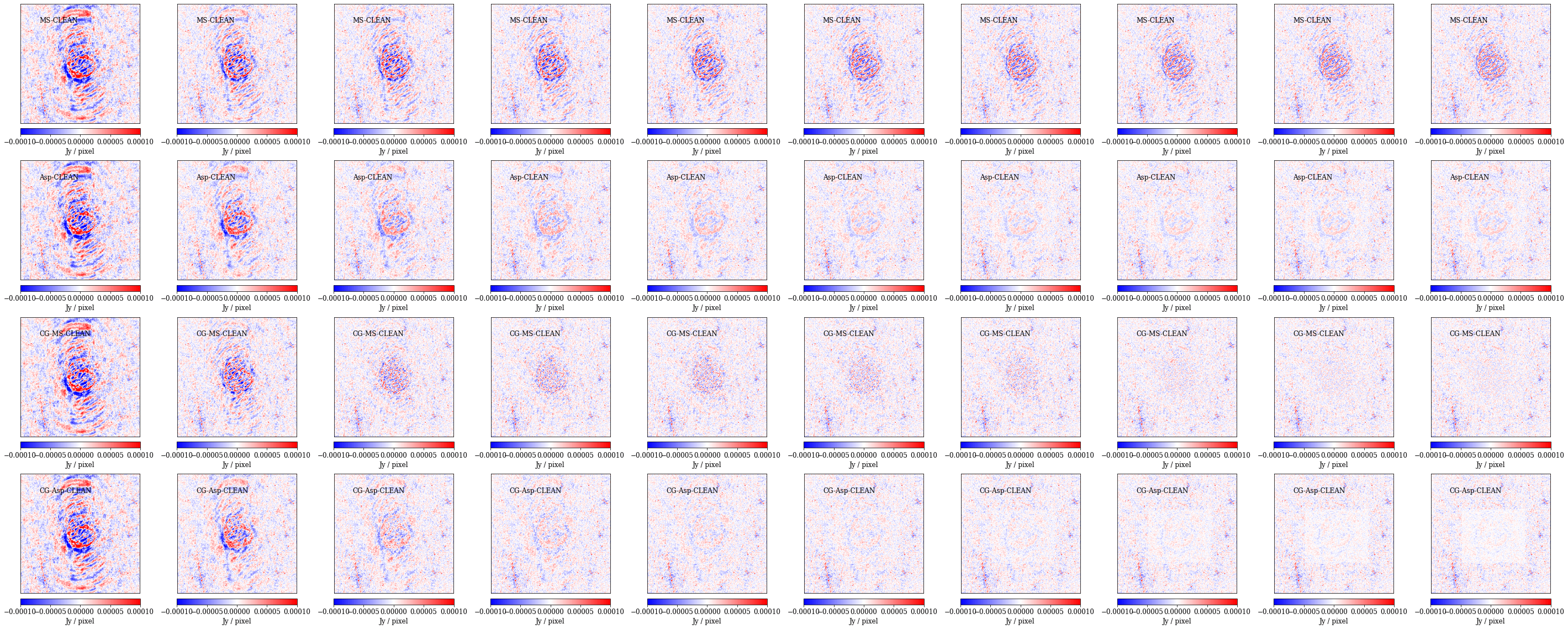}
    \caption{Same as Fig. \ref{fig:cygA_asp}, but for G055.7+3.4.}
    \label{fig:g55_asp}
\end{figure*}

\begin{figure*}
    \centering
    \includegraphics[width=\textwidth]{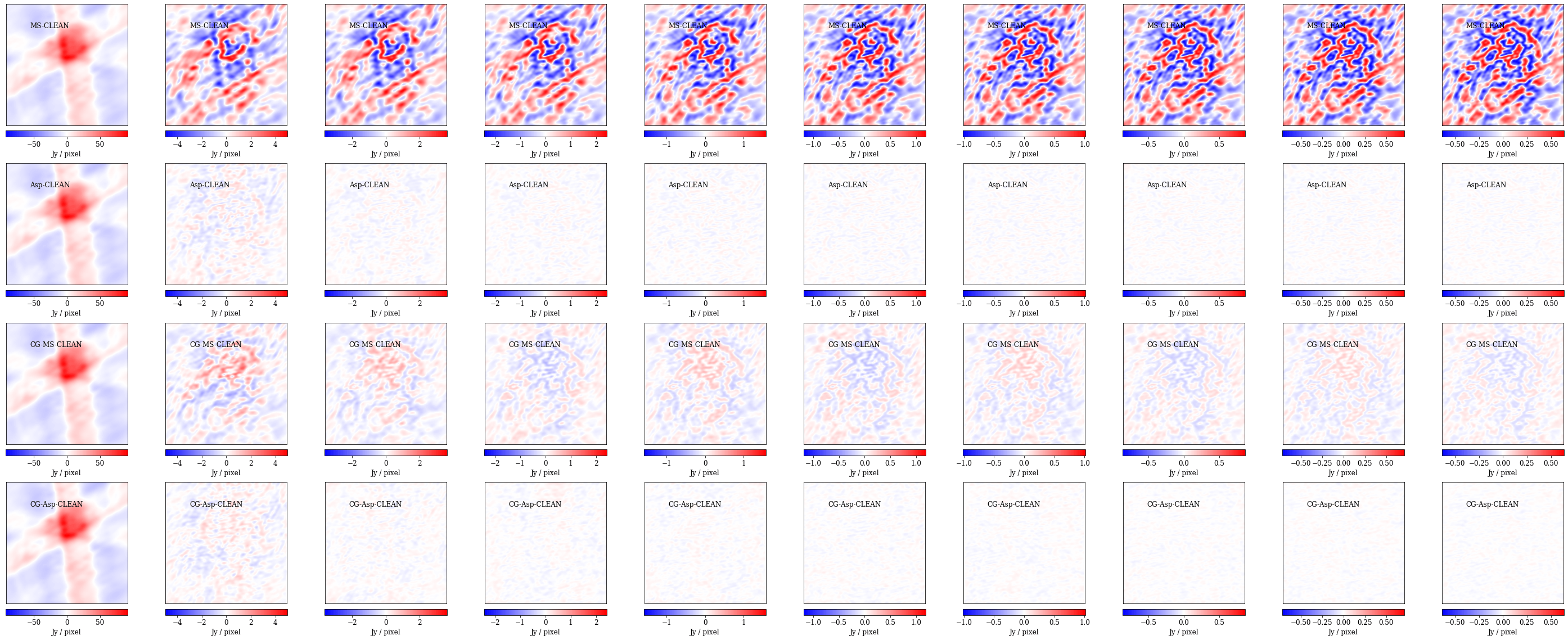}
    \caption{Same as Fig. \ref{fig:cygA_asp}, but for M31.}
    \label{fig:m31_asp}
\end{figure*}

In Fig. \ref{fig:cygA_asp}, Fig. \ref{fig:g55_asp} and Fig. \ref{fig:m31_asp}, we compare the performance improvement by CG-CLEAN stemming from a faster major loop scheme in comparison to the effect of the complementary approach to achieve this from the minor loop. It has been argued that in situations in which the numerica cost of the gridding dominates the numerical cost of the deconvolution and application of the Fast Fourier Transform, more sophisticated deconvolution techniques which allow to CLEAN deeper in every minor cycle may accelerate the overall procedure since there are eventually fewer major loop iterations necessary as a consequence \citep[e.g.][]{Bhatnagar2025}. The most advanced deconvolution algorithm available from standard software packages and commissioned for wide applications may be Asp-CLEAN \citep{Bhatnagar2004, Hsieh2021}. To that end, we compare MS-CLEAN in the traditional Cotton-Schwab cycle (first row), and CG-CLEAN calling MS-CLEAN in the minor loop (third row) with Asp-CLEAN in Cotton-Schwab cycles (second row), and finally combining both, CG-CLEAN calling Asp-CLEAN during the minor loops. While the improvements offered by Asp-CLEAN over MS-CLEAN are remarkable, the acceleration in the convergence speed achieved by the CG scheme over classical CLEAN is comparable, in some cases even better, than the improvements introduced by switching from MS-CLEAN to Asp-CLEAN for the minor loop. The overall best performance can be observed when both approaches are combined (fourth row), albeit the level of improvement is smaller. We can explain this by saturating performance when reaching the noise level. This may be explicitly visible from the G055.7+3.4 residuals for which the simple rectangular mask gets visible, indicating the start of overfitting.

\subsection{More challenging data regimes}
Finally, we apply the best performing algorithm developed in this manuscript, CG-CLEAN, to more challenging data regimes, extending over the simple narrow-band regime discussed until now. We do that in two directions. First we discuss performance in a low resource setting with VLBI observations by the VLBA. Second, we discuss extensions into the wideband regime.

\begin{figure*}
    \centering
    \includegraphics[width=\textwidth]{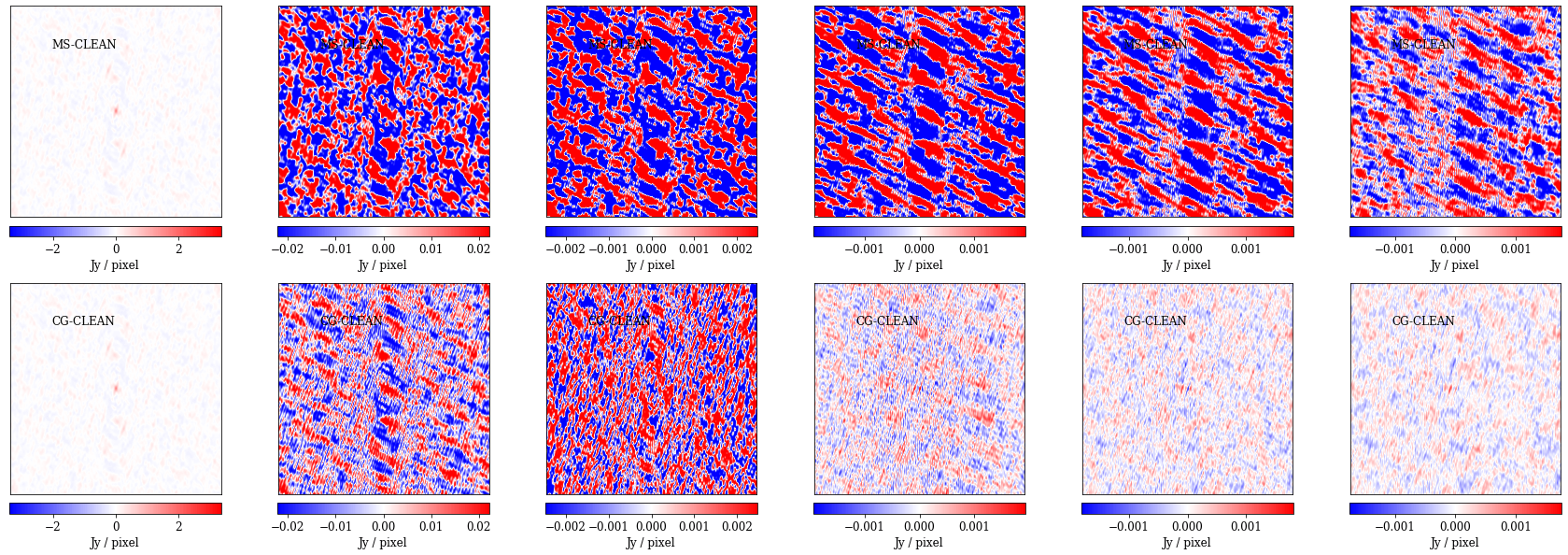}
    \caption{Residual as a function of increasing major loop iterations (from left to right) for MOJAVE observations of 3C120 for CLEAN (top row) and CG-CLEAN (bottom row).}
    \label{fig:mojave}
\end{figure*}

\begin{figure*}
    \centering
    \includegraphics[width=\textwidth]{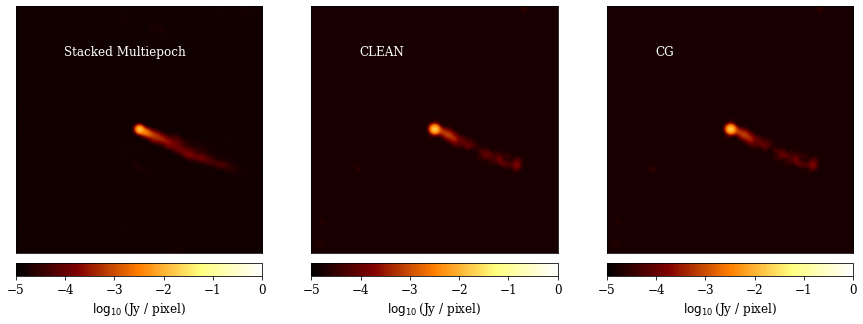}
    \caption{The stacked image of 3C120 (stacked from 30 years of data-taking, left panel) and our reconstructions of a single epoch with MS-CLEAN (middle panel), and CG-CLEAN (right panel).}
    \label{fig:mojave_models}
\end{figure*}

\begin{figure*}
    \centering
    \includegraphics[width=\textwidth]{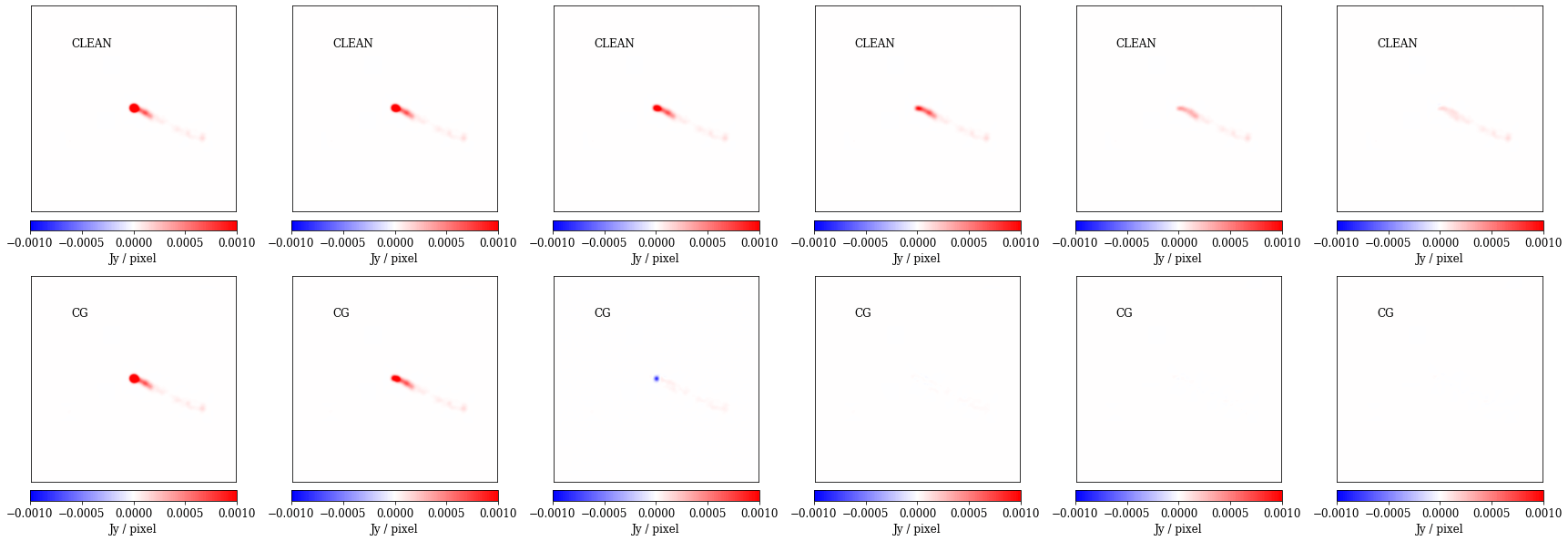}
    \caption{Difference between current model and final model (after the last major loop iteration) as a function of increasing major loop iterations (from left to right) for MOJAVE single epoch observations of 3C120 for CLEAN (top row) and CG-CLEAN (bottom row).}
    \label{fig:mojave_models_iter}
\end{figure*}

As in previous subsections, we show the residual with increasing major loop iteration in Fig. \ref{fig:mojave}, the final models in Fig. \ref{fig:mojave_models}, and the difference between the final model and the recovered model after each major loop iteration in Fig. \ref{fig:mojave_models_iter}. Since MOJAVE is a monitoring program \citep{Lister2018}, we can also compare the reconstructions from a single epoch to reconstructions obtained by stacking images obtained over a time of $\sim 20$ years. By stacking, usually fainter features are highlighted that are challenging to identify in images of individual epochs. However, interpretation of the stacked image may also be misled due to the intrinsic variability of the source.

\begin{figure*}[t!]
    \centering
    \begin{subfigure}[b]{0.7\textwidth}
        \centering
        \includegraphics[width=\textwidth]{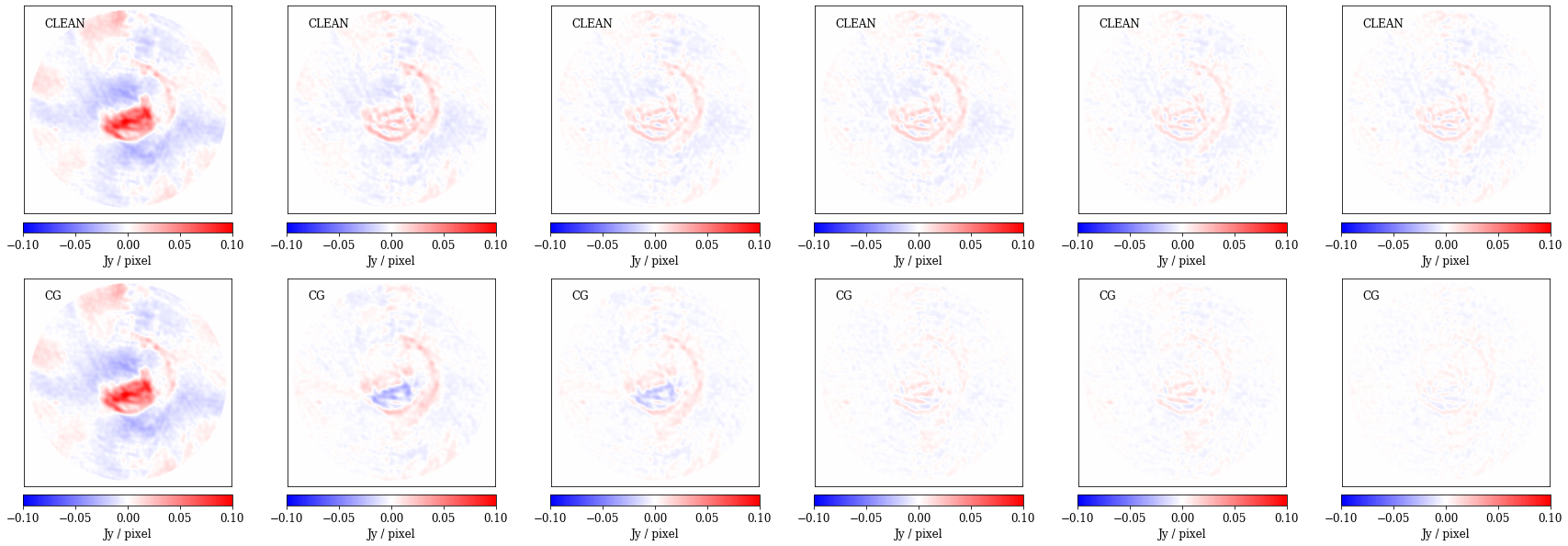}
        \caption{Channel 0}
    \end{subfigure}
    \\
    \begin{subfigure}[b]{0.7\textwidth}
        \centering
        \includegraphics[width=\textwidth]{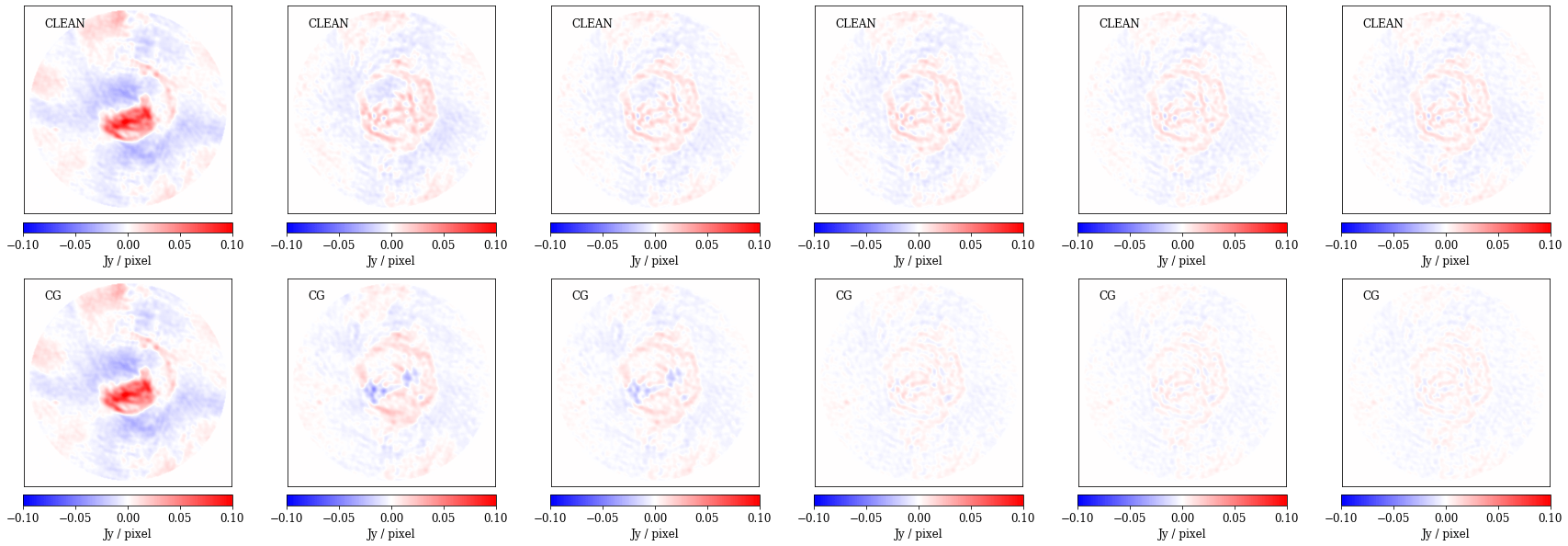}
        \caption{Channel 5}
    \end{subfigure}
    \\
    \begin{subfigure}[b]{0.7\textwidth}
        \centering
        \includegraphics[width=\textwidth]{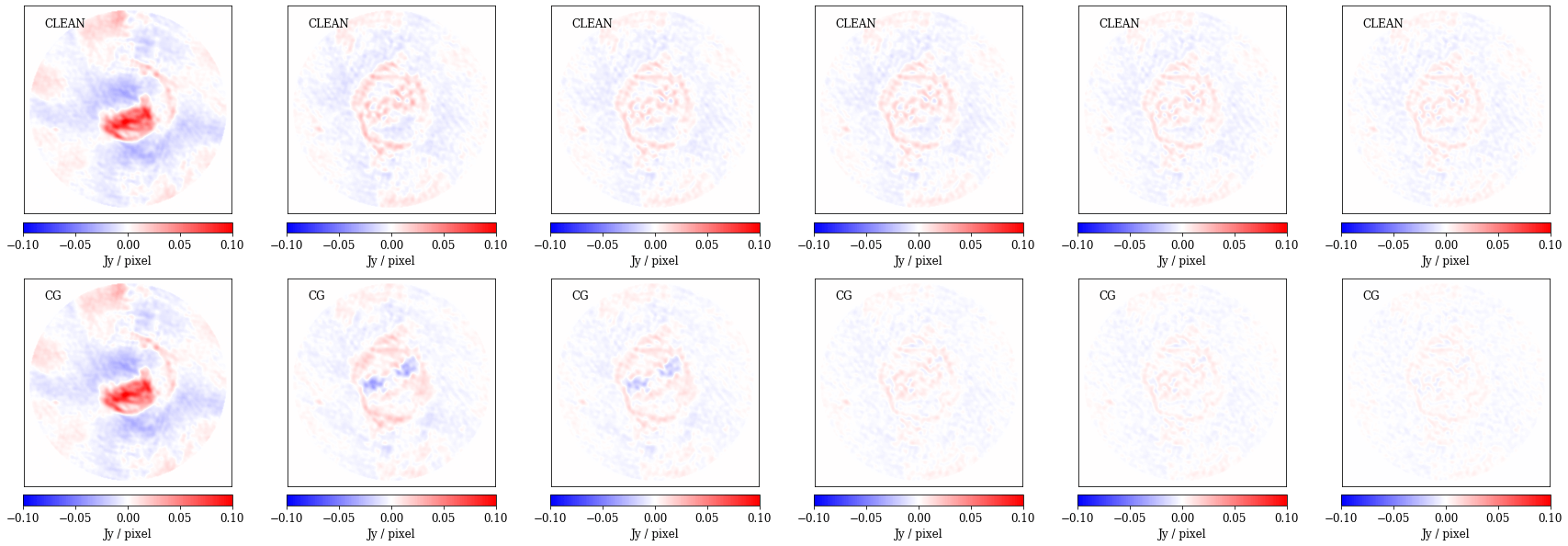}
        \caption{Channel 10}
    \end{subfigure}
    \\
    \begin{subfigure}[b]{0.7\textwidth}
        \centering
        \includegraphics[width=\textwidth]{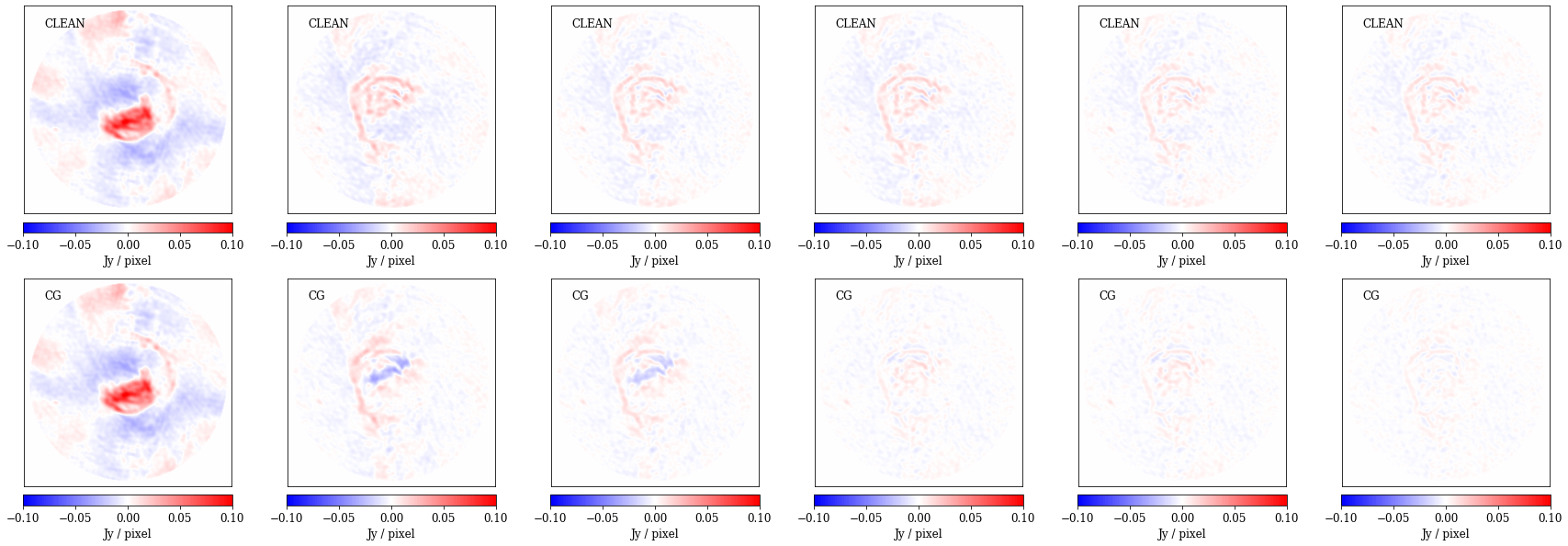}
        \caption{Channel 15}
    \end{subfigure}
    \\
    \begin{subfigure}[b]{0.7\textwidth}
        \centering
        \includegraphics[width=\textwidth]{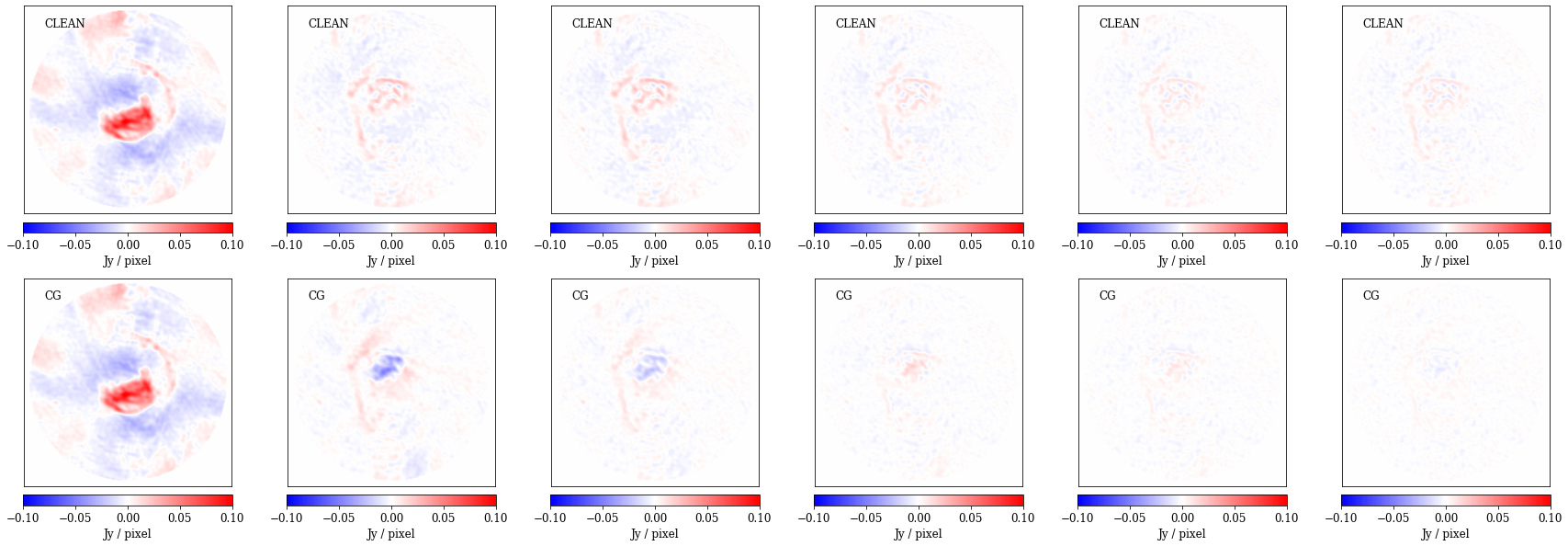}
        \caption{Channel 20}
    \end{subfigure}
    \caption{Residual as a function of increasing major loop iterations (from left to right) for NGC 628 with the VLA for CLEAN (top rows in every subfigure) and CG-CLEAN (bottom row in every subfigure). The subfigures show different spectral channels.}
    \label{fig:spectral_residuals}
\end{figure*}

\begin{figure*}[t!]
    \centering
    \begin{subfigure}[b]{0.4\textwidth}
        \centering
        \includegraphics[width=\textwidth]{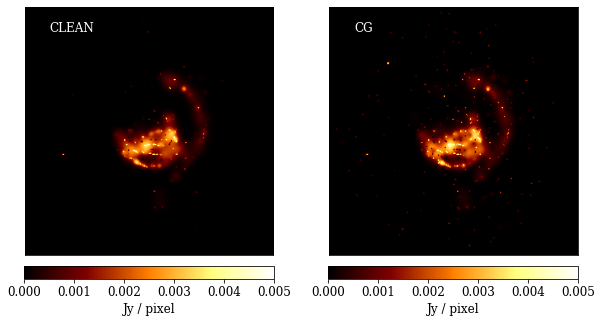}
        \caption{Channel 0}
    \end{subfigure}
    \\
    \begin{subfigure}[b]{0.4\textwidth}
        \centering
        \includegraphics[width=\textwidth]{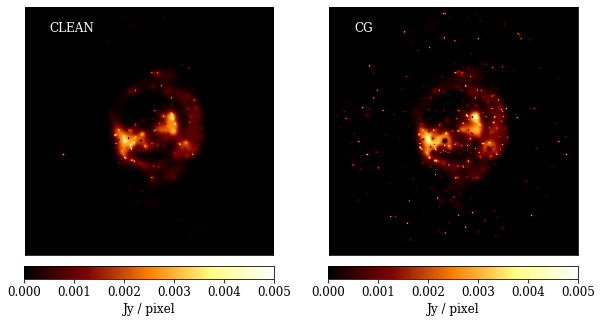}
        \caption{Channel 5}
    \end{subfigure}
    \\
    \begin{subfigure}[b]{0.4\textwidth}
        \centering
        \includegraphics[width=\textwidth]{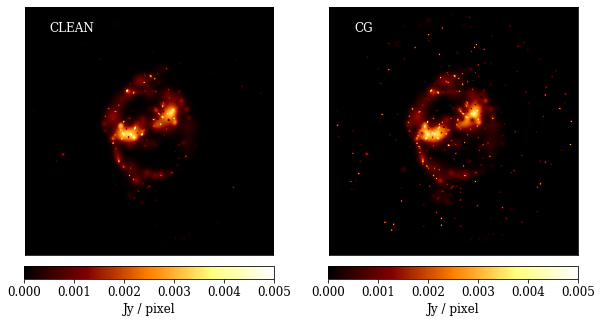}
        \caption{Channel 10}
    \end{subfigure}
    \\
    \begin{subfigure}[b]{0.4\textwidth}
        \centering
        \includegraphics[width=\textwidth]{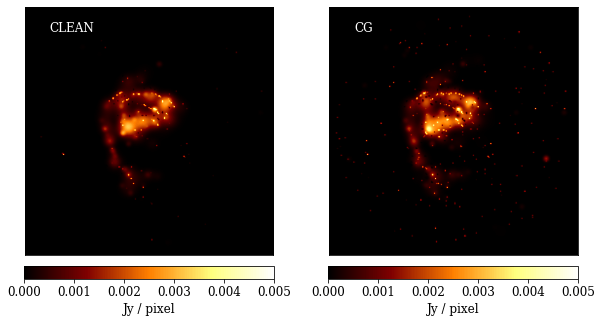}
        \caption{Channel 15}
    \end{subfigure}
    \\
    \begin{subfigure}[b]{0.4\textwidth}
        \centering
        \includegraphics[width=\textwidth]{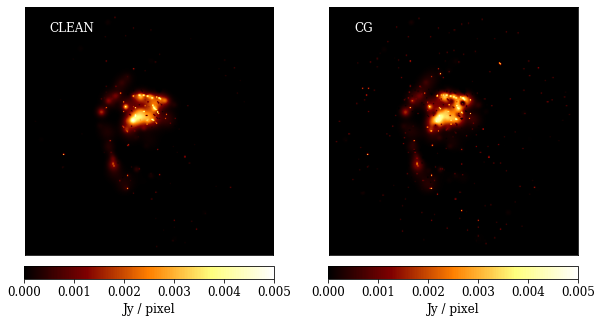}
        \caption{Channel 20}
    \end{subfigure}
    \caption{Recovered models with CLEAN (left column) and CG-CLEAN (right column) for for NGC 628 observations with the VLA. The subfigures show different spectral channels.}
    \label{fig:spectral_models}
\end{figure*}

\begin{figure*}[t!]
    \centering
    \begin{subfigure}[b]{0.7\textwidth}
        \centering
        \includegraphics[width=\textwidth]{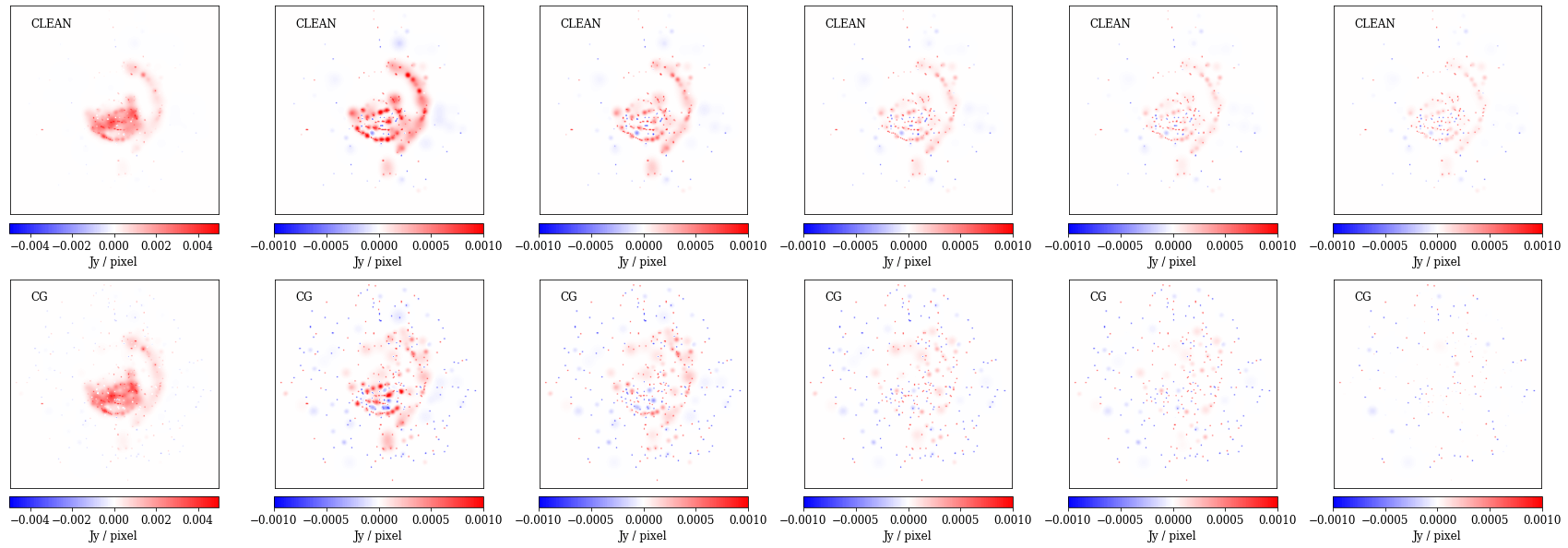}
        \caption{Channel 0}
    \end{subfigure}
    \\
    \begin{subfigure}[b]{0.7\textwidth}
        \centering
        \includegraphics[width=\textwidth]{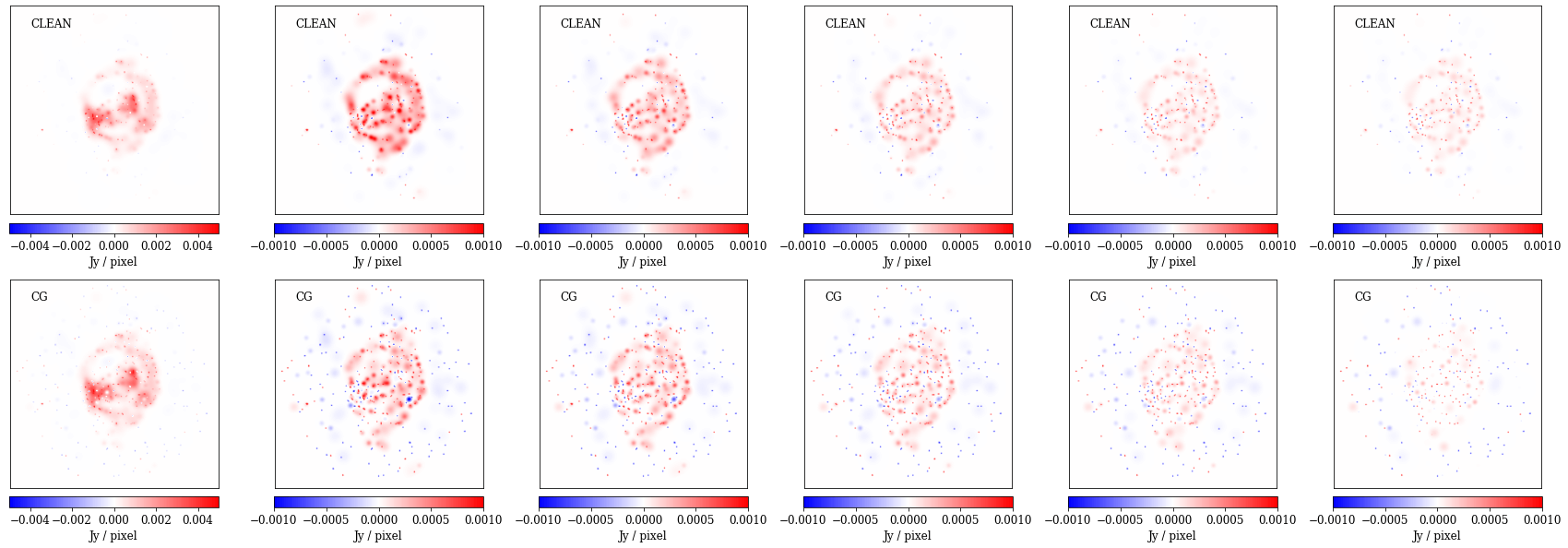}
        \caption{Channel 5}
    \end{subfigure}
    \\
    \begin{subfigure}[b]{0.7\textwidth}
        \centering
        \includegraphics[width=\textwidth]{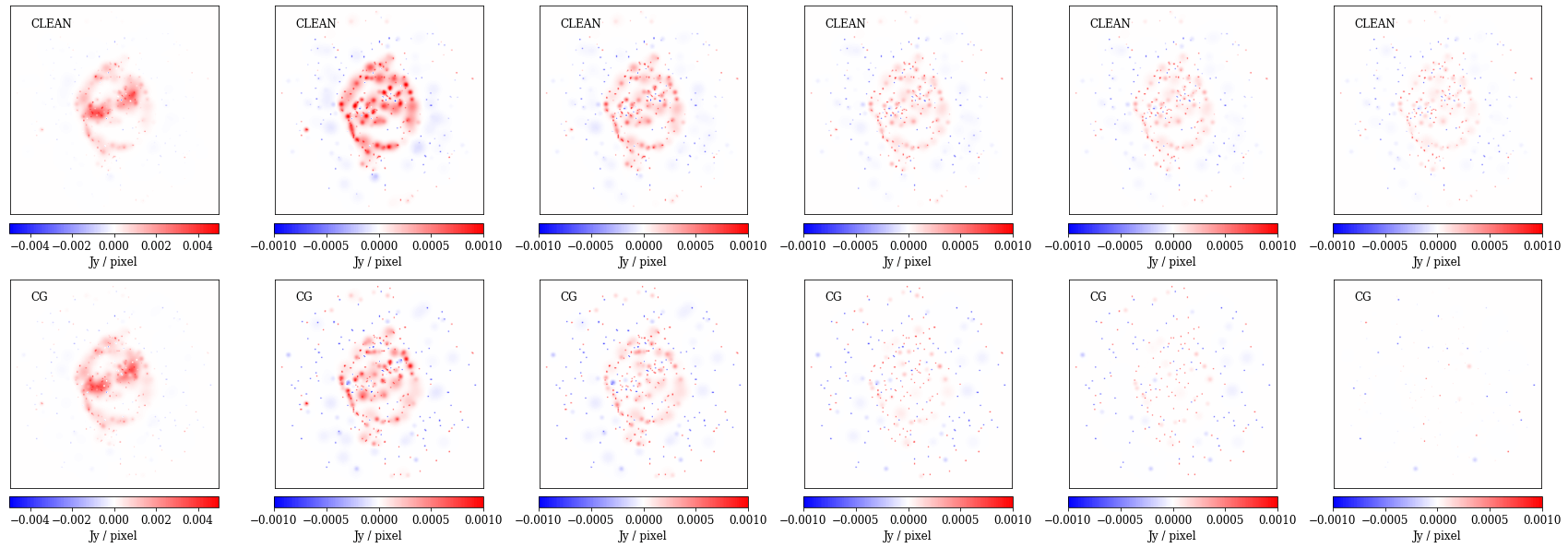}
        \caption{Channel 10}
    \end{subfigure}
    \\
    \begin{subfigure}[b]{0.7\textwidth}
        \centering
        \includegraphics[width=\textwidth]{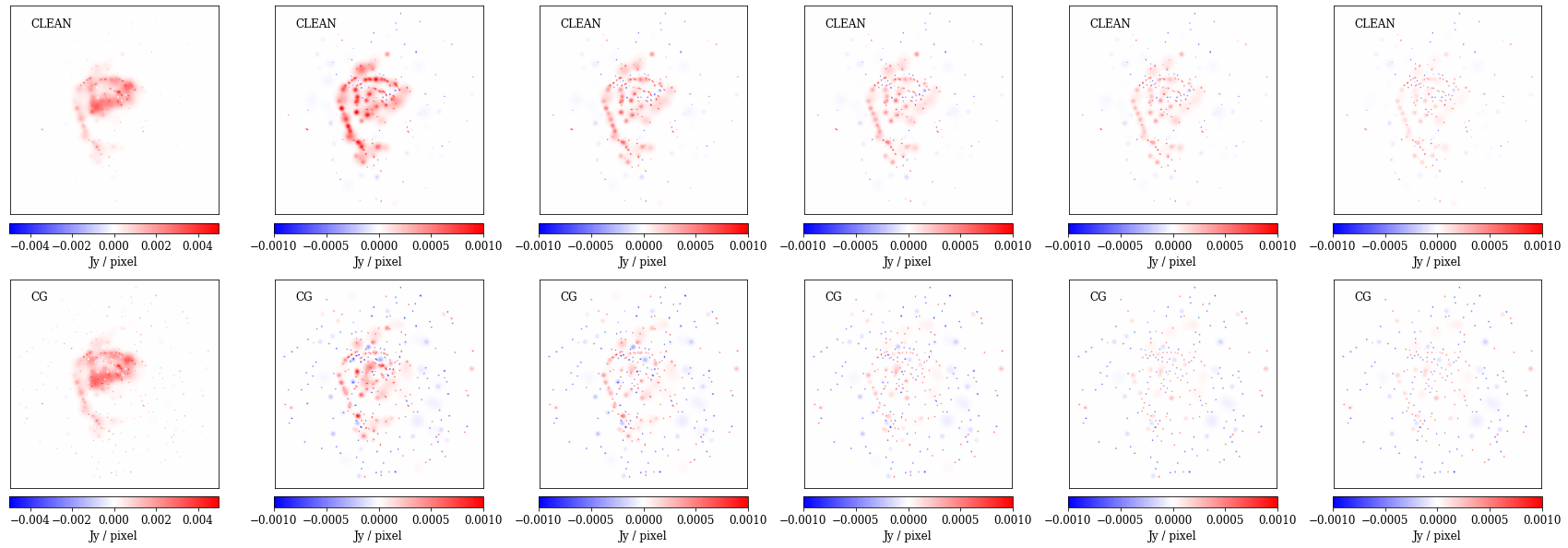}
        \caption{Channel 15}
    \end{subfigure}
    \\
    \begin{subfigure}[b]{0.7\textwidth}
        \centering
        \includegraphics[width=\textwidth]{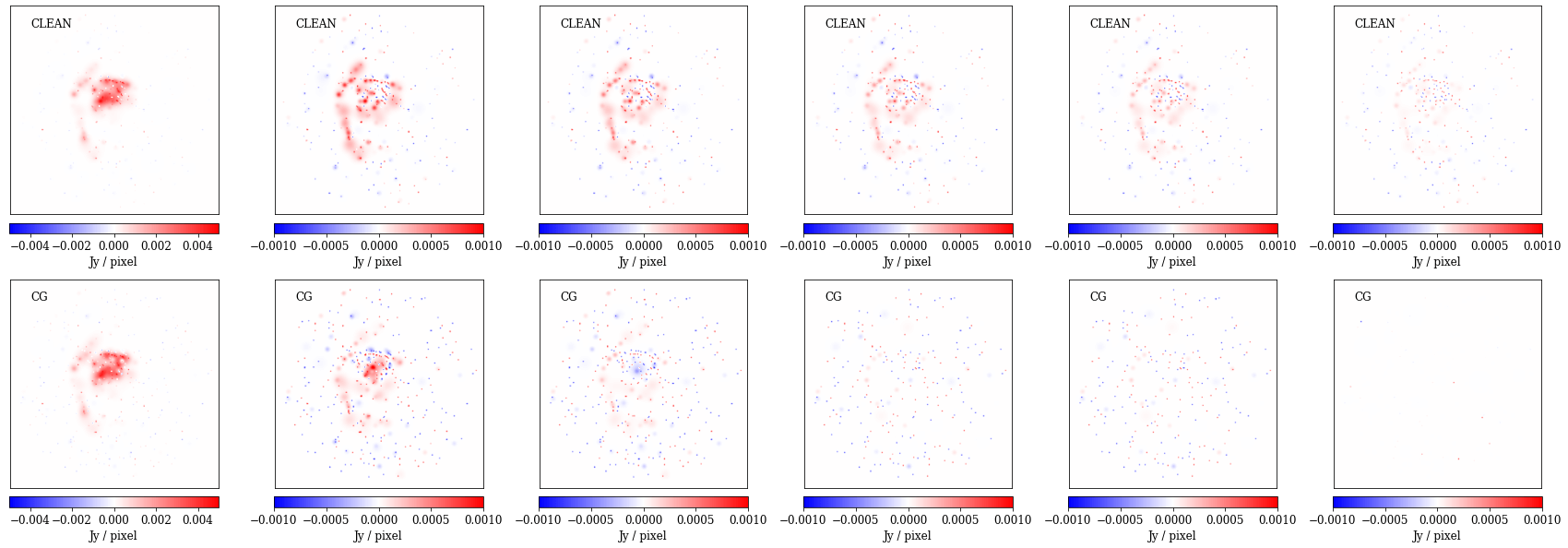}
        \caption{Channel 20}
    \end{subfigure}
    \caption{Difference between current model and final model (after the last major loop iteration) as a function of increasing major loop iterations (from left to right) for NGC 628 with the VLA for CLEAN (top rows in every subfigure) and CG-CLEAN (bottom row in every subfigure). The subfigures show different spectral channels.}
    \label{fig:spectral_models_iters}
\end{figure*}

Finally, we apply the CG-CLEAN algorithm to a spectral reconstruction. As usual, we show the residuals in Fig. \ref{fig:spectral_residuals}, the final models in Fig. \ref{fig:spectral_models} and difference between the recovered model and final models are shown in Fig. \ref{fig:spectral_models_iters}. For the reconstruction, almost exactly the same code is used as in the continuum fitting case: We only need to change the \textit{specmode} keyword in the tclean calls to the minor loop and major loop from \textit{mfs} (the default option for continuum imaging with only one output image channel), to \textit{cube} (the default option with CASA for spectral line imaging with more than one output image channel. The reconstruction has been done in 100 spectral channels representing a widths of $2\,\mathrm{km/s}$. Here, in these figures we show five exemplary channels of them. We note that we discuss narrow-band spectral data here only, and leave a study multiterm continuum imaging for a consecutive work.

These figures demonstrate that the findings that we reported in synthetic data and narrow-band observations also extrapolate to the spectral regime, and VLBI settings: CG-CLEAN produces smaller and noise-like residuals in fewer iterations, results in a final model that is compatible to the model obtained by traditional CLEAN; but needs fewer major loop iterations (and hence is significantly faster) to obtain this result.

We note that the performance margin offered by CG-CLEAN over traditional CLEAN varies for the different channels, most notably it narrows in channel 5. We attribute this effect to limitations in the minor loop (shared by both traditional CLEAN and CG-CLEAN) in representing diffuse emission around the compact core, rather than to a specific issue of CG-CLEAN. In both cases, the minor loop fits scattered delta components across the field instead of assigning the missing flux to a few large-scale structures, causing deconvolution accuracy to saturate. Since CG-CLEAN mainly accelerates convergence across major loops, its benefit diminishes when minor-loop convergence stalls. This behavior is general: CG-CLEAN’s gains are ultimately limited by the underlying calibration and data quality. The specific, described issue is well known and could be mitigated by masking (manual or automated) or by further tuning the scale-bias and loop-gain parameters.

Our result that CG-CLEAN is expanding well into more challenging data regimes is not surprising to us since CG-CLEAN does not represent a particularly more aggressive approach to imaging. The calls to \textit{tclean} to compute the major loop and the minor loop remain the same as for classical Cotton-Schwab CLEAN. This has two important consequences: First, we could, as has been done here, validate the algorithm pretty easily in all scenarios supported by \textit{CASA}, including spectral, widefield, VLBI, polarimetry, mosaics or a combination of them with little extra effort. Second, the heuristics and implementation which was found, validated and commissioned to be robust for most interferometric observations is available to us, especially throughout deconvolution. We have observed that multiscalar stopping rules and scalebiases especially have prevented the overfitting of structures that may be projected into the residual. A full validation and commissioning on all observational data that a radio-interferometer may need to support however is beyond the scope of this manuscript.

\subsection{Amount of momentum memory}
We have demonstrated in the previous subsections that we can accelerate the classical CLEAN approach in some scenarios by adding a momentum in the subtraction of the residuals, or by orthogonalization of the search directions. In fact, referring back to the algorithm presented in Alg. \ref{tab: momentum}, the memory of the momentum of the previous residual update has one free parameter associated to it: the amount of momentum carried to the next gradient update step $\mu$. The parameter $\mu$ shares some similarities with the classical minor loop gain: It regulates the step-size of the CLEANing procedure. $\mu$ is however not equivalent to the loop gain, since $\mu$ affects the step-size of the major loop and the traditional loop gain (which needs to be specified for Momentum-CLEAN as well) the step-size of the minor loop. The most straightforward illustration of the effect of momentum memory (regulated by $\mu$) in the traditional picture would be a loop gain that is varying pixel by pixel in the image, i.e. we allow for steeper component subtraction in regions that have been found to carry significant emission in earlier iterations of the minor loop.

\begin{figure}
    \centering
    \includegraphics[width=\linewidth]{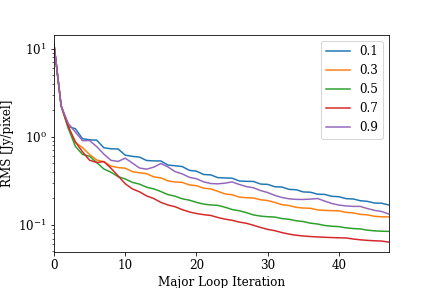}
    \caption{RMS as a function of major loop iteration for M31 obtained with Momentum-CLEAN with different values for $\mu$.}
    \label{fig:rms_momentum}
\end{figure}

In this subsection, we use the reconstruction of M31 to study the impact of this parameter in the reconstructions. First of all, we would like to mention that it is anticipated that $0 \leq \mu \leq 1$. $\mu \leq 0$ would rescind numerical operations done in the previous step. $\mu \geq 1$ would overdo the correction done in previous steps, and leads to pixels that are switching between positive and negative values between iterations. In Fig. \ref{fig:rms_momentum}, we show the norm of the residual as a function of the major loop iteration for different values of $\mu$ for M31. We see the best performance at $\mu = 0.7$, close to the value $\mu = 0.5$ chosen in the previous subsection. The performance worsens again for bigger values. Moreover, after closer inspection of the red and pink line ($\mu=0.7$ and $\mu=0.9$), the convergence curve shows an oscillatory behavior for which the reconstruction even goes worse for some iterations. The same effect is visible in Fig. \ref{fig:rms} for Cygnus A. We attribute this oscillating behavior to the consequence of switching pixels (i.e. some pixels get over-cleaned) when we add too much momentum mentioned above. 

\section{Conclusions} \label{sec:conclusions}
In this manuscript, we examine the classical Cotton-Schwab CLEAN cycle. Despite the numerous alternative methods proposed over the years, CLEAN remains the de facto standard, primarily due to its robustness and computational efficiency. We offer a novel perspective on the algorithm by interpreting it as a Newton minimization technique. Although this interpretation does not constitute a mathematical equivalence, since several key aspects of the CLEAN algorithm, such as its matching pursuit nature, are not captured in this formulation, and the fundamental assumptions of Newton's method are violated, this analogy serves as an illustrative framework.

In particular, we demonstrate that the division of CLEAN into major and minor loops can be understood in terms of the individual steps involved in computing the gradient (major loop) and inverting the Hessian (minor loop). Building on this analogy, we propose a method for identifying new CLEAN variants by the inverse direction of arguments: we begin with promising minimization algorithms and express them in terms of the robust and numerically efficient major and minor loop iterations, by replacing the gradient evolution with the major loop and the inversion of the Hessian with the minor loop. We apply this framework to several minimization algorithms, including implicit gradient descent, conjugate gradient descent (applied to the normal equation), and heavy-ball (Nesterov) acceleration techniques.

It is important to note that the methodology presented here could, in principle, be extended to a broader range of algorithms with relative ease. The chosen algorithms are not meant to represent a comprehensive list, but rather a set of examples to demonstrate the approach. Additionally, we acknowledge that much of the ongoing research into novel minimization techniques is focused on the development of quasi-Newton and frozen Newton methods, which are unlikely to yield significant improvements in the context of the scope of this manuscript, given that the Hessian is typically known in this domain and we explicitly aim to apply it here through the minor loop, rather than approximate.

We benchmark two variants of CLEAN: CG-CLEAN (conjugate gradients interpreted as CLEAN) and Momentum-CLEAN (heavy-ball momentum acceleration applied to CLEAN), against the traditional CLEAN algorithm using multiple test observation. We applied a hierarchically more challenging validation ladder to our procedures, consisting of data from the CASA tutorial on 3C391, synthetic and real narrow-band data, and a spectral and VLBI data reduction. We observe in all of these examples that CG-CLEAN is producing smaller residuals than CLEAN in fewer iterations, obtains compatible, or even better reconstructions than CLEAN, and is obtaining these results with fewer major loop iterations, the driving cost during data processing. We found similar trends for Momentum-CLEAN in some of the test data. We interpret these findings as CG-CLEAN achieving significantly faster convergence, requiring fewer major loop iterations. Additionally, CG-CLEAN cleans to deeper levels than the original CLEAN, providing improvements in dynamic range that are comparable to, or even exceed, those offered by state-of-the-art minor loop algorithms such as Asp-CLEAN over traditional MS-CLEAN. The best performance is achieved by combining the CG-CLEAN framework in the major loop with Asp-CLEAN operating in the minor loop, allowing the algorithm to reach the noise level after just a few major loop iterations. The claim of faster convergence towards an agreed on final model is supported by an analysis on synthetic data, and a comparison to the ground truth where CG-CLEAN shows higher and faster rising PSNR than traditional techniques. These improvements are realized with only a minimal increase in computational cost, and in a framework that can be straightforwardly expanded to more challenging scenarios, including VLBI, wideband, widefield or mosaicing by reusing the robust implementations already available through CASA. We have exemplary tested the first two regimes, and report the same degree of acceleration.

This work contributes to efforts aimed at scaling data processing techniques for the large data volumes expected from next-generation radio interferometers, such as the ngVLA, SKA, and DSA2000. Given that the primary computational cost for these arrays will be the gridding process, any approach that reduces the number of major loop iterations is highly valuable. While the long-term future for data processing may not lie within CLEAN, but a variety of more sophisticated imaging procedures already proposed and in development, the short-term and mid-term pipelines may benefit by how straightforwardly especially CG-CLEAN is built upon existing pipelines. Moreover, choice of algorithms for future arrays also necessitates the further development of CLEAN methods, at least as a benchmark for the limits of traditional radio interferometric imaging.

Finally, we emphasize that the algorithms discussed in this manuscript are conceptually straightforward and built upon robust principles and existing software foundations. As a result, adapting these algorithms into state-of-the-art data processing software should involve minimal challenges.

\begin{acknowledgements}
The implementation of this work was performed in the software package \textit{CASA} \citep{CASA2022}. In its earlier development stage of this manuscript, we however made use of software tools providing less monolithic, more modular and simple low-level access to radio-astronomic data. In particular, we would like to acknowledge \textit{LibRA} \footnote{\url{https://github.com/ARDG-NRAO/LibRA}}, and \textit{MrBeam} \footnote{\url{https://github.com/hmuellergoe/mrbeam}} \citep{Mueller2022, Mueller2024a}. Special thanks goes to Preshanth Jagganathan for his support with the installation and handling of \textit{LibRA}. This research was supported through the Jansky fellowship program of the National Radio Astronomy Observatory. NRAO is a facility of the National Science Foundation operated under cooperative agreement by Associated Universities, Inc.. Furthermore, H.M. acknowledges support by the M2FINDERS project which has received funding from the European Research Council (ERC) under the European Union’s Horizon 2020 Research and Innovation Program (grant agreement No 101018682).
\end{acknowledgements}
 
\bibliographystyle{aa}
\bibliography{lib}{}

\appendix

\section{Proof of orthogonality} \label{app: orthogonality}

We will now show by induction over the number of major loop iteration $k$ that the search directions are orthogonal onto each other with respect to the beam, i.e. we show $p_i^+ B p_j = 0$ for indices $i \neq j$ and $I_k^+ p_i = 0$ for $i < k$. For the induction beginning, we compute $I_1^+ p_0$. It is: 
\[
I_1^+ p_0 = I_0^+ p_0 - \alpha_0 p_0^+ B p_0 = 0,
\]
by the definition of $\alpha_0$. Moreover, we have:
\[
p_1^+ B p_0 = z_1^+ B p_0 + \beta_0 p_0^+ B p_0 = 0,
\]
by the definition of $\beta_0$. Now we consider the induction step, $k \rightarrow k+1$. It is for $i < k+1$:
\[
I_{k+1}^+ p_i = I_k^+ p_i - \alpha_k p_k^+ B p_i.
\]
If $i = k$, then the right hand side vanishes due to the definition of $\alpha_k$. If $i < k$, then both terms vanish due to the induction assumption. Similarly, we have:
\[
p_{k+1}^+ B p_i = z_{k+1}^+ B p_i + \beta_k p_{k}^+ B p_i.
\]
If $i = k$, then the right hand side vanishes due to definition of $\beta_k$. If $i<k$, the second term on the right hand side vanishes due to the induction assumption. For the first term, note that $B z_{k+1} \approx I_{k+1}$, so the first term vanishes as well.

We would like to note that these properties only hold exactly in the absence of gridding and calibration errors. In practice however, one would typically need to stop the minor loop earlier based on manually enforced stopping criteria before the noise-level is reached and potentially manipulate the dataset itself (e.g. self-calibration or flagging). As a consequence the relation $B z_{k} = I_{k}$ holds only approximately. Moreover, the relation $I_{k+1} = I_{k} - \alpha_k B p_k$ also holds only approximately in the presence of gridding errors. Both are needed to ensure exact orthogonality of the search directions $p_k$. Application in practice may benefit from restarting CG iterations. Nevertheless, the arguments listed above show that at least directly following search directions $p_k, p_{k+1}$ are orthogonal on each other.

We note that the necessity to restart the CG iterations is not conceptually different from the wide-spread application of the algorithm outside the interferometry context. It is a fairly common scenario due to departures from orthogonality which accumulate due to numerical errors. However, specifically in radio interferometry, there are additional sources of errors such as gridding and calibration.

There are two trivial corollaries to these identities. Since $z_{k} \in \mathrm{span}[p_{k-1},p_k]$, we also have $I_k^+ z_i = 0$ for all $i < k$. Similarly, we get for $i < k$, $I_i^+ z_k = z_i^+ B z_k = I_k^+ z_i = 0$. We will make use of these identities below.

Note that for $k \geq 1$: 
\[
I_k^+ p_k = I_k^+ z_k + \beta_{k-1} I_k^+ p_{k-1} = I_k^+ z_k.
\]
Hence, we also get:
\begin{align}
    \alpha_k = \frac{I_k^{res,+}  z_k}{p_k^+ B p_k}. \label{eq: alpha_alt}
\end{align}
Moreover, it is:
\[
\alpha_k z_{k+1}^+ B p_k = -z_{k+1}^+ I_{k+1}. 
\]
Together, we get:
\begin{align}
    \beta_k = \frac{I_{k+1}^+ z_{k+1}}{I_k^+ z_{k}} \label{eq: beta_alt}
\end{align}
Eq. \eqref{eq: alpha_alt} and Eq. \eqref{eq: beta_alt} define equivalent, alternative definitions for the step-size parameters $\alpha$ and $\beta$.

\section{Figures of our reconstruction results}

\end{document}